%% file: main.tex
\documentclass[10pt,aps,pra,twocolumn,superscriptaddress,nofootinbib,floatfix,longbibliography]{revtex4-2}

\usepackage{placeins}
\usepackage{fix-cm}

\usepackage{silence} 
\usepackage{microtype}
\usepackage{graphicx}
\usepackage[english]{babel}
\usepackage{amsmath,amssymb}
\usepackage{bm}
\usepackage[colorlinks=true,allcolors=blue]{hyperref}
\usepackage[capitalise,nameinlink]{cleveref}
\usepackage{braket}
\usepackage{multirow}
\usepackage{booktabs}
\usepackage{orcidlink}
\usepackage[table, dvipsnames]{xcolor}
\usepackage{tikz}
\usepackage{ragged2e}

\usetikzlibrary{quantikz2,calc}

\graphicspath{{figures/}}
\newcommand{\ii}{\mathrm{i}\mkern1mu} 

\makeatletter
\clo@superscriptaddress
\makeatother

\begin{document}

\title{Quantum walk-based optimisation for capacitated vehicle routing \texorpdfstring{\\}{ } with homogeneous and heterogeneous fleets}
\author{Edric Matwiejew}
\email{edric.matwiejew@pawsey.org.au}
\affiliation{Pawsey Supercomputing Research Centre, Perth, WA 6152, Australia}
\author{Aidan Smith}
\affiliation{Centre for Quantum Information, Simulation and Algorithms, The University of Western Australia, Perth, WA 6009, Australia}
\author{Callum Neill}
\affiliation{Centre for Quantum Information, Simulation and Algorithms, The University of Western Australia, Perth, WA 6009, Australia}
\author{Paulo Santos}
\affiliation{PrioriAnalytica, Adelaide, SA 5000, Australia}
\author{Ugo Varetto}
\affiliation{Pawsey Supercomputing Research Centre, Perth, WA 6152, Australia}
\author{Jingbo Wang}
\affiliation{Centre for Quantum Information, Simulation and Algorithms, The University of Western Australia, Perth, WA 6009, Australia}
\date{\today}

\begin{abstract}
The capacitated vehicle routing problem (CVRP) is an appealing candidate for quantum optimisation due to its combinatorial complexity and practical importance. However, the problem's constrained search space poses a challenge for such quantum algorithms. We introduce a quantum walk-based optimisation algorithm (QWOA) for the CVRP with homogeneous or heterogeneous vehicle fleets, addressing this challenge through a continuous-time quantum walk over a product space that coincides with combinatorial structures intrinsic to the CVRP solution space. Relative to the prior QWOA-based formulation, this approach reduces the per-layer gate complexity from $\mathcal{O}(n^{3}\log n)$ to $\mathcal{O}(n^{2}\log n)$ and supports a circuit parameterisation schedule generated by a fixed number of classical parameters. Exact state-vector simulation on instances with up to $n=8$ customers and $K=3$ vehicles demonstrates improved convergence to low-cost solutions using markedly fewer objective function evaluations, with the advantage broadening as problem size increases. These results identify structured product-space walks as a promising tool for optimisation over constrained combinatorial spaces.
\end{abstract}

\maketitle

\section{Introduction}
\label{sec:introduction}

The Capacitated Vehicle Routing Problem (CVRP) is a core challenge in combinatorial optimisation and logistics~\cite{dantzig1959truck}. It seeks a minimum-cost collection of routes for a fleet of capacity-constrained vehicles that must collectively serve all customers from a central depot. As a constrained generalisation of the Travelling Salesman Problem, the CVRP is NP-hard~\cite{TOTH2002487}, and the number of feasible solutions grows super-exponentially with the number of customers. The problem has wide-ranging applications in parcel delivery, waste collection, transit scheduling, and inventory distribution, and even marginal improvements to routing solutions can yield substantial cost savings over time~\cite{Braekers2016}.

Classical approaches to the CVRP span exact, heuristic, and metaheuristic methods. Exact methods such as branch-and-bound and branch-and-cut can certify optimality, but are typically restricted to relatively small instances~\cite{Laporte1992}. For larger problems, practical solvers usually rely on heuristics and metaheuristics, including the Clarke--Wright savings algorithm, genetic algorithms, and tabu search, which trade guarantees of optimality for scalability~\cite{Braekers2016,TOTH2002487}. This combination of practical importance and computational hardness makes the CVRP a natural test case for quantum optimisation methods~\cite{Osaba2022}.

The Quantum Approximate Optimisation Algorithm (QAOA)~\cite{farhi2014quantum} is a variational hybrid quantum-classical algorithm for combinatorial optimisation. It generates approximate solutions by applying an alternating sequence of two parameterised quantum operations over $p$ layers, comprising a \emph{phase-separation unitary}, which encodes the problem solution space as phases on the computational basis states, and a \emph{mixer unitary}, which transfers amplitude between basis states producing solution-cost dependent interference. A classical outer loop optimises the resulting $2p$ parameters to increase the probability of identifying optimal or near-optimal solutions. The Quantum Walk-based Optimisation Algorithm (QWOA)~\cite{Marsh2019,marsh_combinatorial_2020} builds on this framework, defining the mixing unitary as a continuous-time quantum walk (CTQW). The walk is generated by the adjacency matrix of a graph over candidate solutions, allowing the quantum state to move selectively between computational basis states connected by the graph's edges. 

An important refinement of this framework is the use of \emph{problem-specific mixing graphs}, whose connectivity reflects the combinatorial structure and constraints of a given problem type~\cite{marsh_combinatorial_2020,matwiejew_quantum_2024}. Tailoring the graph in this way yields correlations between inter-vertex distances and variations in solution costs, improving the convergence behaviour and scalability of the ansatz~\cite{matwiejew_quantum_2024,bennett_non-variational_2024}. A further development is the ``non-variational'' QWOA~\cite{bennett_non-variational_2024}, which introduced a parameterisation scheme in which three coupled parameters generate all $2p$ ansatz parameters for any number of layers $p$. The parameter count is therefore independent of both problem size and circuit depth, dramatically reducing the overhead of the classical optimisation loop even at high layer counts.

Prior work on quantum algorithms for vehicle routing has explored both quantum annealing~\cite{irie2019quantumannealingvehiclerouting,harikrishnafkumar2020quantum,borowski2020new,SYRICHAS201752,10.3389/fict.2019.00013} and gate-based methods~\cite{bennett_quantum_2021,Osaba2022,Fitzek2024}. QAOA formulations for the CVRP typically embed the routing constraints as penalty terms in the computational basis and apply a standard \emph{transverse-field} mixing unitary~\cite{Osaba2022,Fitzek2024}. Alternatively, a previously proposed QWOA for the CVRP employs an \emph{indexing unitary} to restrict its search to a subspace of valid routings~\cite{bennett_quantum_2021}. However, these algorithms aim to satisfy problem constraints~\cite{Fitzek2024,bennett_quantum_2021}, rather than exploit the combinatorial structure of the feasible solution space. 

In this work, we develop an improved QWOA for the CVRP with both \emph{homogeneous} and \emph{heterogeneous} fleets. It employs a problem-specific mixing graph together with a product-space encoding of the solution space, where each solution is represented as an unordered assignment of customers to vehicles and a permutation of these unordered assignments. The resulting ansatz restricts amplitude to a space of valid vehicle routings without the indexing unitary used in the earlier indexed-QWOA formulation, thereby reducing the per-layer gate complexity by a polynomial factor. Numerical simulations using the constant-count-parameter schedule of~\cite{bennett_non-variational_2024} indicate that the new approach achieves improved convergence to low-cost solutions and a substantial reduction in classical optimisation overhead.

The remainder of the paper is as follows. \Cref{sec:background} introduces the CVRP and QWOA framework. \Cref{sec:schema} develops the solution-space encoding, problem-specific mixing graphs, and the required quantum circuits. \Cref{sec:numerical} presents numerical methods and results. We discuss our findings in \cref{sec:discussion} and conclude in \cref{sec:conclusion}.

\section{Background}
\label{sec:background}

\subsection{The capacitated vehicle routing problem}
\label{sec:cvrp}

Here, we introduce the CVRP formulation considered in this work~\cite{laporte2009,bennett_quantum_2021}. The locations of a problem instance are assigned to a set of nodes $\{0,1,\dots,n\}$, with $0$ denoting the depot and $1,\dots,n$ the customers. The fleet consists of $K\in\mathbb{N}$ vehicles. Each vehicle $k$ has capacity $Q_k > 0$ and a cost modifier $\alpha_k > 0$ that scales the travel costs on its route. Each customer $i$ has demand $d_i \ge 0$ (with $d_0=0$ for the depot), and the cost of travelling from node $i$ to node $j$ is $c_{ij}\ge 0$, with $c_{ii}=0$ and $c_{ij} \neq c_{ji}$ permitted. 

A solution $x=(r_1,\dots,r_K)$ consists of $K$ routes, where each route $r_k=(r_{k,1},\dots,r_{k,n_k})$ contains the sequence of customers served by vehicle $k$, and $n_k\ge 0$ is the number of customers on that route. Each vehicle departs from and returns to the depot, so with the convention $r_{k,0}=r_{k,n_k+1}=0$, the total travel cost is
\begin{equation}
\label{eq:cvrp-cost}
f(x) = \sum_{k=1}^{K} \alpha_k \Bigl[\;\sum_{j=0}^{n_k} c_{r_{k,j},\,r_{k,j+1}}\;\Bigr].
\end{equation}
A solution is \emph{feasible} if each customer is visited exactly once by one vehicle and each route satisfies the capacity constraint $\sum_{j=1}^{n_k} d_{r_{k,j}} \le Q_k$. We place no constraint on the total cost of the per-vehicle routes. The set of feasible solutions is denoted as $\Omega_{\mathrm{feas}}$. 

\begin{figure}[t]
    \centering
  \includegraphics[width=\columnwidth]{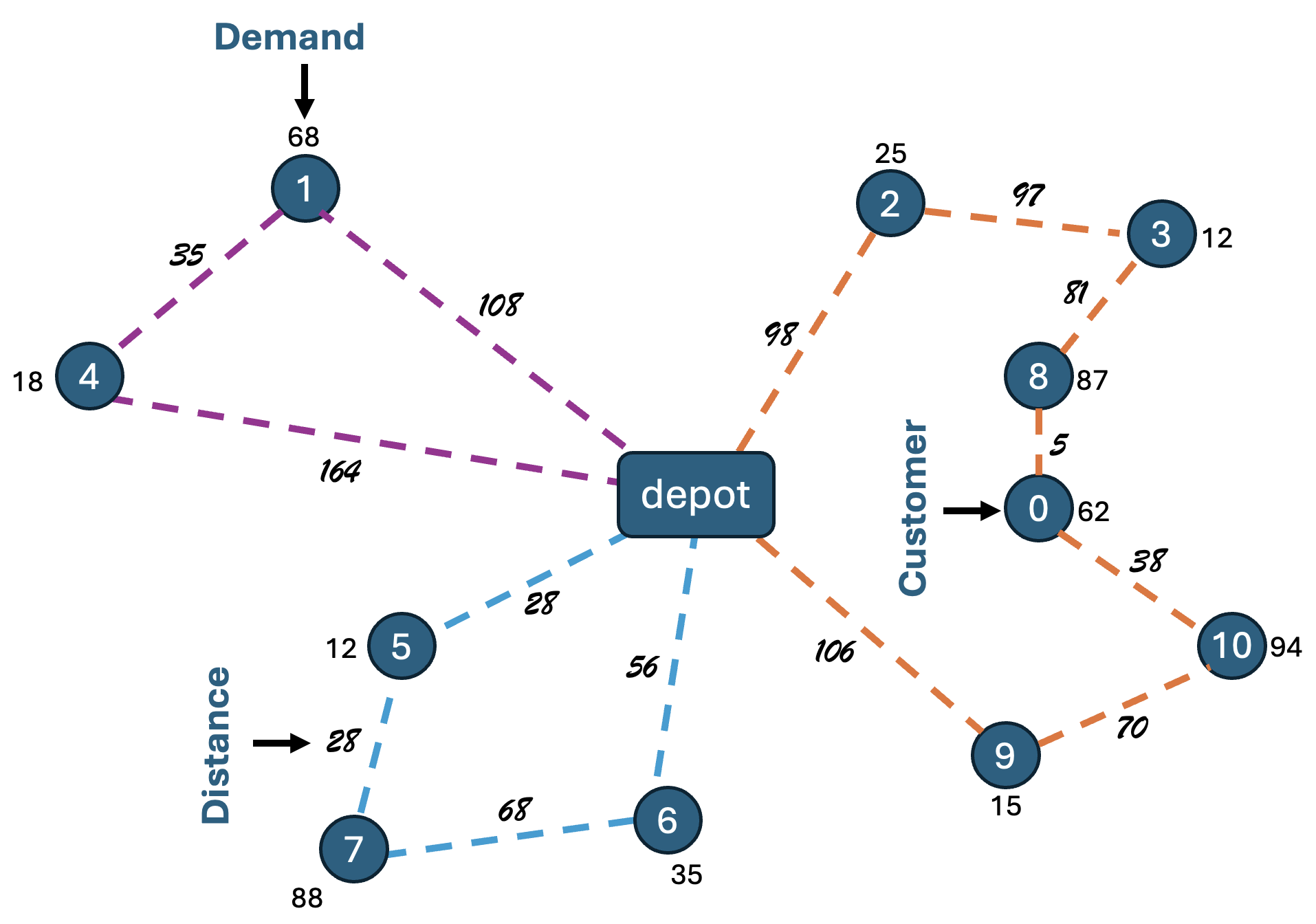}
    \caption{Schematic of a CVRP instance. The central depot dispatches vehicles along distinct routes (coloured paths) to serve customers (numbered nodes). Each node is annotated with its demand, and each edge with its travel cost. Every route must begin and end at the depot, and the total demand on each route must not exceed the vehicle capacity.}
    \label{fig:cvrp_schematic}
\end{figure}

Altogether, the goal of the CVRP is the minimisation of the cost function
\begin{equation}
\label{eq:cvrp-minimisation}
  x^\star \in \arg\min_{x\in\Omega_{\mathrm{feas}}} f(x),
  \qquad
  f^\star = f(x^\star).
\end{equation}
In lieu of the globally optimal solution $x^\star$, sufficiently low-cost solutions are also accepted. We illustrate an example CVRP in \cref{fig:cvrp_schematic}. When all vehicles are identical (i.e., $Q_k = Q$ and $\alpha_k = 1$ for all $k$), the problem is the \emph{homogeneous CVRP}~\cite{Braekers2016}. When vehicles differ in capacity or operating cost, the problem is the \emph{heterogeneous CVRP} (also known as the heterogeneous fleet vehicle routing problem~\cite{TothVigo2014}).

Enforcing capacity constraints directly within a QAOA or QWOA framework is computationally costly as it introduces a solution-value-dependent connectivity into the mixing unitary graph. Instead, routes that exceed the vehicle capacity incur an additive penalty
\begin{equation}
\label{eq:penalised-cost}
    f_\lambda(x)  = f(x) + \lambda \sum_{k=1}^{K} \Bigl[\;\sum_{j=1}^{n_k} d_{r_{k,j}} - Q_k\;\Bigr]_+,
\end{equation}
where $[y]_+ = \max\{y,0\}$ and $\lambda\ge 0$ sets the penalty strength~\cite{bennett_quantum_2021}. We refer to the set of such solutions, which encode the routing structure with only this ``soft'' enforcement of the capacity constraint, as the \emph{canonical solution space}, denoting it $\Omega_{\mathrm{het}}$ for the heterogeneous CVRP and $\Omega_{\mathrm{hom}}$ for the homogeneous case. 

For the heterogeneous CVRP, each solution in $\Omega_{\mathrm{het}}$ is specified by a permutation of the $n$ customers together with a weak composition of $n$ into $K$ route lengths~\cite{Stanley2011EC1}, giving a total of
\begin{equation}
\label{eq:solution_count}
|\Omega_{\mathrm{het}}| = n!\binom{n+K-1}{K-1} = \frac{(n+K-1)!}{(K-1)!}
\end{equation}
canonical solutions. The homogeneous CVRP has a smaller canonical solution space:
\begin{equation}
\label{eq:hom_solution_count}
|\Omega_{\mathrm{hom}}| = \sum_{k=1}^{K} \operatorname{Lah}(n,k),
  \qquad
  \operatorname{Lah}(n,k)=\binom{n{-}1}{k{-}1}\frac{n!}{k!},
\end{equation}
where $\operatorname{Lah}(n,k)$ is an unsigned Lah number counting the partitions of $\{1,\dots,n\}$ into $k$ nonempty totally ordered subsets~\cite{bennett_quantum_2021,petkovsek2007lah}.

\subsection{The quantum walk-based optimisation algorithm}
\label{sec:qwoa}

We now describe the QWOA framework~\cite{Marsh2019,marsh_combinatorial_2020}. The starting point is a mapping of the canonical solution space $\Omega$ to a Hilbert space $\mathcal H$,
\[\Omega\longrightarrow\mathcal{H},\qquad x\mapsto\ket{x},\]
and realisation of the classical cost function $f:\Omega\to\mathbb{R}$ as the diagonal cost operator
\begin{equation}
\label{eq:cost-op}
\mathcal{C}=\sum_{x\in \Omega} f(x)\,\ket{x}\!\bra{x}.
\end{equation}

The QWOA then defines an ansatz comprised of layers of two interleaved unitaries. A \emph{phase-separation unitary}
\begin{equation}
\label{eq:phase-sep}
  U_{\mathcal{C}}(\gamma) = e^{-\ii \gamma \mathcal{C}},
\end{equation}
applies the phase rotation $\gamma f(x)$ to each $\ket{x}$, with variational parameter $\gamma\in\mathbb R$.

Next, a \emph{mixing unitary} redistributes amplitude within a solution subspace $\mathcal{S} \subseteq \Omega$ via a CTQW: a time evolution whose generator is defined by the adjacency matrix of an undirected graph $G=(V, E)$ with vertex set $V$ and edge set $E$. In practice, the mixing unitary can comprise several commuting \emph{sub-mixers}, each defined by a CTQW on a different graph $G_j=(V_j, E_j)$~\cite{matwiejew_quantum_2024}. The corresponding CTQW generators are
\begin{equation}
\label{eq:generator}
\mathcal{G}_j=\sum_{(x,x')\in E(G_j)} w^{(j)}_{xx'}\bigl(\ket{x}\!\bra{x'}+\ket{x'}\!\bra{x}\bigr),
\end{equation}
where $w^{(j)}_{xx'}$ are real-valued edge weights. The composite mixing unitary then factorises as
\begin{equation}
\label{eq:composite-mixer}
  U_{\mathcal{G}}(t) = \prod_{j} e^{-\ii t\, \mathcal{G}_j},
\end{equation}
where $t\ge 0$ is the shared walk time.

The typical choice for the sub-mixer graphs $G_j$ is to connect solutions that differ by the smallest non-zero Hamming distance~\cite{matwiejew_quantum_2024,bennett_non-variational_2024}. Because such a minimal change typically produces a relatively small cost difference, this produces a correspondence between inter-vertex distance and variation in solution cost, improving the extent to which the phase-separation and mixer unitaries can together reinforce transfer towards low-cost states~\cite{matwiejew_quantum_2024,bennett_non-variational_2024}. In comparison, mixing graphs that do not coincide with problem structure, such as the complete graph, have a mixing operation that is equivalent to an unstructured search and are therefore upper-bounded by the quadratic convergence of Grover's Algorithm~\cite{xie2025grover_mixer_bound}.

Starting from an efficiently prepared initial state $\ket{\psi_0}$ (typically a superposition over an invariant subspace of the mixing unitary), the $p$-layer QWOA ansatz is
\begin{equation}
\label{eq:qwoa-ansatz}
\ket{\psi(\vec{\gamma},\vec{t})}
   = \prod_{\ell=1}^{p} U_{\mathcal{G}}(t_\ell)\,U_{\mathcal{C}}(\gamma_\ell)\,\ket{\psi_0},
\end{equation}
where $\vec{\gamma}=(\gamma_1,\dots,\gamma_p)$ and $\vec{t}=(t_1,\dots,t_p)$.

A classical outer loop minimises the variational objective function
\begin{equation}
\label{eq:objective-function}
\langle \mathcal C\rangle_p(\vec{\gamma},\vec{t})
  = \bra{\psi(\vec{\gamma},\vec{t})}\mathcal C\ket{\psi(\vec{\gamma},\vec{t})}
  = \sum_{x\in\Omega} f(x)\,P_{\vec{\gamma},\vec{t}}(x),
\end{equation}
where $P_{\vec{\gamma},\vec{t}}(x)=|\braket{x}{\psi(\vec{\gamma},\vec{t})}|^2$.

\subsubsection{Indexed formulation}
In its original formulation, the QWOA addresses constrained combinatorial optimisation and non-bijective solution-space embeddings by introducing an \emph{indexing unitary} $U_\#$ that maps a representative subset of encoded solution states onto an indexed subspace $\mathcal H_\#=\mathrm{span}\{\ket{0},\ldots,\ket{|\Omega|-1}\}$, where $\Omega_{\mathrm{can}}$ denotes the canonical solution space~\cite{bennett_quantum_2021,marsh_combinatorial_2020}. Given a bijection $id:\Omega_{\mathrm{can}}\rightarrow\{0,\ldots,|\Omega_{\mathrm{can}}|-1\}$, the unitary $U_\#$ is implemented via the compute-swap-uncompute pattern~\cite{marsh_combinatorial_2020}
\begin{equation}
  \ket{x}\ket{0} \rightarrow \ket{x}\ket{id(x)} \rightarrow \ket{id(x)}\ket{x} \rightarrow \ket{id(x)}\ket{0}.
\end{equation}
In this formulation, $\ket{\psi_0}$ is prepared as a uniform superposition over $\mathcal H_\#$, and the mixing unitary is defined by a complete graph that fully connects the indexed states~\cite{marsh_combinatorial_2020,bennett_quantum_2021}.

\subsubsection{Three-parameter schedule}
\label{sec:three_parameter_schedule}

The parameter schedule of the \emph{non-variational QWOA}~\cite{bennett_non-variational_2024}, reduces the $2p$ variational parameters of the QWOA to at most three. For $p>1$, the layer-wise values are
\begin{equation}
\label{eq:gamma_k}
\gamma_\ell=\frac{\gamma}{\sigma}\Bigl[\beta+(1-\beta)\frac{\ell-1}{p-1}\Bigr],
\end{equation}
and
\begin{equation}
\label{eq:t_k}
t_\ell=t\Bigl[1+(\beta-1)\frac{\ell-1}{p-1}\Bigr],
\end{equation}
with $\gamma_1=\gamma/\sigma$ and $t_1=t$ when $p=1$. The taper parameter $\beta\in[0,1]$ controls how the phase-separation and mixing weights vary across layers, and $\sigma$ is the standard deviation of the diagonal of the cost operator, which in practice can be estimated by classical random sampling over a modest number of solutions~\cite{bennett_non-variational_2024}. The linear profile and the normalisation of 
$\gamma$ by $\sigma$, are motivated in~\cite{bennett_non-variational_2024}.

\section{Product-Space QWOA}
\label{sec:schema}

\subsection{Solution space encoding}
\label{sec:cvrp_encoding}

We represent CVRP solutions using two combinatorial components, each associated with a multi-qubit register. The \emph{permutation register} $\mathcal H_\pi$ has basis $\{\ket{\pi}:\pi\in S_n\}$, where $S_n$ denotes the symmetric group on $n$ elements and $\pi$ is a permutation of the $n$ customer locations. It has dimension $|\mathcal H_\pi| = n!$ and is encoded in $n$ blocks of $\lceil \log_2 n \rceil$ qubits. The \emph{assignment register} $\mathcal H_a$ has basis $\{\ket{a}:a\in\{1,\dots,K\}^n\}$, where $a=(a_1,\dots,a_n)$ and $a_j=k$ indicates that the customer at position $j$ of $\pi$ is served by vehicle $k$. It has dimension $|\mathcal H_a| = K^n$ and requires $n$ blocks of $\lceil \log_2 K \rceil$ qubits. In the remainder of this work, we refer to the QWOA built on this encoding as the \emph{product-space QWOA} (PS-QWOA). Together, the combined \emph{encoding Hilbert space} of the PS-QWOA is
\begin{equation}
  \label{eq:encoding_space_dimension}
  \left|\mathcal H_\pi\otimes\mathcal H_a\right| = n!\,K^n.
\end{equation}

Routes are obtained by grouping consecutive positions of permutation $\pi$ that share the same vehicle assignment in $a$. For example, with $n=4$ customers and $K=3$ vehicles, $\pi=(3,2,1,4)$ and $a=(1,3,2,2)$ encode the routes $(0,3,0)$, $(0,1,4,0)$, and $(0,2,0)$ for vehicles 1, 2, and 3, respectively, each starting and ending at the depot. This product-space encoding satisfies the CVRP constraint that each customer is assigned to exactly one vehicle as each customer appears exactly once in any permutation $\pi$.

\subsection{Unitaries}
\label{sec:walk_generators}

We begin with a state-preparation unitary $U_{\mathrm{prep}} = U_{\mathrm{prep},\pi} \otimes U_{\mathrm{prep},a} $ which prepares the uniform initial state
\begin{equation}
\label{eq:initial_state_full}
\begin{aligned}
    U_{\mathrm{prep}}\ket{0}_\pi\ket{0}_a
      &= \ket{\psi_0}
       = \ket{\psi_\pi}\otimes\ket{\psi_a} \\
      &= \frac{1}{\sqrt{n!\,K^n}}
         \sum_{\pi \in S_n}\sum_{a \in \{1,\dots,K\}^n}\ket{\pi, a}.
\end{aligned}
\end{equation}

Specialising the cost operator, \cref{eq:cost-op}, to the penalised cost function $f_\lambda$ of \cref{eq:penalised-cost} gives
\begin{equation}
\label{eq:cost-op-cvrp}
\mathcal{C} = \sum_{\pi\in S_n}\sum_{a \in \{1,\dots,K\}^n} f_\lambda(\pi,a)\,\ket{\pi,a}\!\bra{\pi,a},
\end{equation}
with the phase-separation unitary $U_{\mathcal{C}}(\gamma)$ as defined in \cref{eq:phase-sep}.

The mixing unitary is composed of two commuting sub-mixers that act on the permutation and assignment registers, respectively. The \emph{permutation sub-mixer} couples states whose permutations differ by a single transposition while sharing the same assignment. Two states $\ket{\pi,a}$ and $\ket{\pi',a}$ are adjacent whenever $\pi'=\pi\circ(i\;j)$ for some pair $i\ne j$, which corresponds to a Hamming distance of $h(\pi,\pi')=2$. The resulting graph is a copy of the transposition Cayley graph of $S_n$ for each fixed $a$ with vertex degrees $\binom{n}{2}$. The corresponding CTQW generator is
\begin{equation}
\label{eq:G_pi}
\begin{aligned}
\mathcal G_\pi
  &= \frac{1}{\binom{n}{2}}\Biggl[
     \sum_{a\in\{1,\dots,K\}^n}
     \sum_{\substack{\{\pi,\pi'\}\subset S_n\\ h(\pi,\pi')=2}} \\
  &\qquad\times
    \bigl(
      \ket{\pi',a}\bra{\pi,a}
      + \ket{\pi,a}\bra{\pi',a}
    \bigr)\Biggr].
\end{aligned}
\end{equation}
The \emph{assignment sub-mixer} couples states whose vehicle assignments differ in exactly one position while sharing the same permutation. Each vertex has degree $n(K{-}1)$. For each fixed $\pi$, this produces a copy of the Hamming graph $H(n,K)$, such that the corresponding CTQW generator is
\begin{equation}
\label{eq:G_a}
\begin{aligned}
\mathcal G_a
  &= \frac{1}{n(K-1)}\Biggl[
     \sum_{\substack{\{a,a'\}\subset\{1,\dots,K\}^n\\ h(a,a')=1}}
     \sum_{\pi\in S_n} \\
  &\qquad\times
    \bigl(
      \ket{\pi,a'}\bra{\pi,a}
      + \ket{\pi,a}\bra{\pi,a'}
    \bigr)\Biggr],
\end{aligned}
\end{equation}
with Hamming distance between $a$ and $a'$ of $h(a,a')$. In both $\mathcal G_\pi$ and $\mathcal G_a$, we divide by the vertex degree so that the two walks have the same total incident edge weight.

Since $\mathcal G_\pi$ and $\mathcal G_a$ act independently on the two registers, they commute and may be applied concurrently. We adopt $t_\pi=t_a=t$, giving a single walk-time parameter per layer that is compatible with the three-parameter linear schedule of~\cref{sec:three_parameter_schedule}. Altogether, the PS-QWOA composite mixing unitary is:
\begin{equation}
\label{eq:mixer}
U_{\mathcal{G}}(t)
   = e^{-\ii t \mathcal G_\pi}\,
     e^{-\ii t \mathcal G_{a}}
   = e^{-\ii t (\mathcal G_\pi + \mathcal G_{a})}.
\end{equation}
The assignment and permutation walk graphs and the resulting structure of $\mathcal G_a + \mathcal G_\pi$ are illustrated in \cref{fig:walk_generator_graphs_n3k2} for $n=3$ and $K=2$. An overview of the PS-QWOA circuit is given in \cref{fig:qwoa_circuit}. 

\input{figures/walk_generator_graphs}

\begin{figure}[tb]
\centering
\begin{quantikz}[row sep=0.25cm, column sep=0.35cm,
                 slice style={draw=black, dashed},
                 every node/.append style={transform shape}, scale=1.00, font=\fontsize{9.2}{8}\selectfont]
\lstick{$\ket{0}^{\otimes n\lceil\log_2 K\rceil}_a$}
  & \gate[1]{U_{\mathrm{prep},a}} \slice{} \qwbundle{}
          & \gate[3]{U_{\mathcal{C}}(\gamma_\ell)}
          & \gate{e^{-\ii t_\ell \mathcal{G}_a}}
          & \qw \slice{$\times\, p$}
          & \meter{} \rstick{$a$} \\
\lstick{$\ket{0}^{\otimes n\lceil\log_2 n\rceil}_\pi$}
    & \gate[1]{U_{\mathrm{prep},\pi}} \slice{} \qwbundle{} &  & \gate{e^{-\ii t_\ell \mathcal{G}_\pi}} & \qw
          & \meter{} \rstick{$\pi$} \\
\lstick{$\ket{0}^{\otimes \mathcal O (b)}$}
    & \qw \qwbundle{} & & \qw & \qw
          & \rstick{$\ket{0}$}
\end{quantikz}
\caption{The PS-QWOA circuit for the CVRP, showing the separability of the state preparation (\cref{eq:initial_state_full}) and mixing unitary (\cref{eq:mixer}) over the assignment $a$ and permutation $\pi$ registers. Each of the $p$ ansatz layers applies the phase-separation unitary (\cref{eq:phase-sep}) to the $a$ and $\pi$ registers, which evaluates the penalised problem cost function (\cref{eq:penalised-cost}) on an $\mathcal{O}(b)$-qubit ancilla register to precision $\mathcal{O}(2^{-b})$, followed by application of the mixing unitary.}
\label{fig:qwoa_circuit}
\end{figure}

\subsection{Properties of the PS-QWOA search space}
\label{sec:cvrp_reduction}

The PS-QWOA encoding is many-to-one for both the homogeneous and heterogeneous CVRP as any permutation of customers that preserves the per-vehicle ordering yields a different $(\pi, a)$ pair that encodes the same solution. If vehicle $k$ serves $n_k$ customers, then the number of degeneracies due to this \emph{interleaving} is
\[
\binom{n}{n_1,\,n_2,\,\dots,\,n_K}. 
\]
This interleaving degeneracy is largest when $n_k \approx n/K$ for all $k$ and equals one when a single vehicle serves all $n$ customers.

However, in the homogeneous case, the initial state, cost function and both walk generators are invariant under global permutation of the vehicle labels by the symmetric group $S_K$. As a result, the dynamics remain in the \emph{effective} subspace $\mathcal H_\pi\otimes\mathcal H_a^{\mathrm{sym}}$, where $\mathcal H_a^{\mathrm{sym}}$ is the $S_K$-symmetric subspace of $\mathcal H_a$. This subspace may be represented by pairing each permutation $\pi\in S_n$ with an equal superposition over all assignments that differ only by a relabelling of the vehicles, giving dimension
\begin{equation}
\label{eq:effective_hom_ps_dimension}
\dim\!\bigl(\mathcal H_\pi \otimes \mathcal H_a^{\mathrm{sym}}\bigr)=n!\,M(n,K),
\end{equation}
\begin{equation}
\label{eq:restricted_bell}
M(n,K)=\sum_{k=1}^{K}\frac{1}{k!}\sum_{j=0}^{k}(-1)^{k-j}\binom{k}{j}j^n,
\end{equation}
where $M(n,K)$ is the number of partitions of $n$ elements into at most $K$ sets~\cite{Stanley2011EC1}. This dimension satisfies $|\Omega_{\mathrm{hom}}| < n!\,M(n,K) < n!\,K^n$, so the PS-QWOA effective subspace is smaller than the full encoding space, but strictly larger than the canonical solution space. A detailed characterisation of the reduced basis, and the corresponding representation of the initial state and walk generators is given in \cref{sec:orbit_derivation}.

In the heterogeneous case, the cost operator $\mathcal{C}$ is not invariant under vehicle relabelling. So, while different pairs $(\pi,a)$ can still encode the same heterogeneous solution, this is due only to the interleaving degeneracy; not, in general, symmetry in the cost operator. As such, the effective Hilbert space given heterogeneous vehicles is the full encoding space of the PS-QWOA.

\subsection{Quantum circuits and complexity}
\label{sec:circuits}

The PS-QWOA solution space encoding requires $n\lceil\log_2 n\rceil$ qubits for the permutation register and $n\lceil\log_2 K\rceil$ qubits for the assignment register, totaling $n(\lceil\log_2 n\rceil + \lceil\log_2 K\rceil)$ data qubits.

The state preparation unitary (see \cref{eq:initial_state_full}) factorises across the two encoding registers. Preparation of $\ket{\psi_\pi}$ on the permutation register as per \cite{Binkowski2025} requires $\mathcal{O}(n^2\log n)$ two-qubit gates and $\mathcal{O}(\log n)$ ancilla qubits. On the assignment register, preparation of $\ket{\psi_a}$ as described in \cite{bennett_non-variational_2024} has gate complexity $\mathcal{O}(nK)$.

Computation of the solution costs is the dominant factor in the complexity of the phase-separation unitary. Under the standard bounded-precision assumption that all input data and intermediate cost-oracle values are polynomially bounded in $n$, the required arithmetic-register width is $b=\mathcal{O}(\log n)$. In this case, the cost operator has gate complexity $\mathcal{O}(n^2 + n\sqrt{n} + nKb)$ for the homogeneous CVRP and $\mathcal{O}(n^2 + n\sqrt{n} + nKb + n\sqrt{K} + nb^2 + K\sqrt{K})$ in the heterogeneous case (see \cref{sec:cost_oracle_details}).

A circuit for the CTQW generated by $\mathcal G_\pi$ is built from block-$\mathrm{SWAP}_{i,j}$ operations between the $n$ permutation register blocks. Because swaps commute only on disjoint block pairs, we use first-order Trotterisation across $\mathcal{O}(n)$ parallel rounds of disjoint pairs with step $\tau=t/\binom{n}{2}$. Each $e^{-\ii\tau\,\mathrm{SWAP}_{i,j}}$ can be implemented exactly with one ancilla and $\mathcal{O}(\log n)$ two-qubit gates~\cite{Nielsen2000,Barenco1995}, giving a total of $\mathcal{O}(n^2\log n)$ two-qubit gates and up to $\lfloor n/2\rfloor$ ancillas when all commuting pairs execute concurrently~\cite{Kivlichan2018}.
The CTQW circuit for $\mathcal G_a$ is given explicitly in~\cite{matwiejew_quantum_2024,bennett_non-variational_2024} and requires $\mathcal{O}(n^2)$ two-qubit gates and $\mathcal{O}(n)$ ancillas. Per-layer mixing cost is therefore dominated by the permutation walk at $\mathcal{O}(n^2 \log n)$ gates.

Since at most $n$ vehicles can serve nonempty routes, worst-case asymptotic scaling occurs at $K=n$. In that case, the PS-QWOA ancilla overhead scales as $\mathcal{O}(nb)$ and two-qubit gate count as $\mathcal{O}(n^2\log n)$ per ansatz layer.

\section{Numerical simulations}
\label{sec:numerical}

\subsection{Methods}
\label{sec:methods}

We compare the PS-QWOA to the previously proposed indexed QWOA (I-QWOA) formulation~\cite{bennett_quantum_2021} on benchmark sets of homogeneous and heterogeneous CVRP instances. In terms of their fundamental unitary dynamics, both QWOA variants share the same problem cost function, but differ in their effective search space and mixing graph. 

\subsubsection{Problem instances}

Each problem instance places $n$ customers and a depot in a two-dimensional plane, with the pairwise costs $c_{ij}$ computed from a sum of the Euclidean distance and small antisymmetric perturbation. Over-capacity routes are penalised linearly in the amount of excess demand using the default rule $\lambda=\lambda_{\text{mean}}$ of \cref{eq:lambda_mean}, where the mean is taken over the off-diagonal costs $c_{ij}$ and $\kappa=\max_k\alpha_k$; this choice is justified in \cref{sec:penalty_selection}. 

Homogeneous instances covered $n\in\{5,6,7,8\}$ with $K=3$ and at $n=8$ with $K=2$, and heterogeneous instances covered $n\in\{5,6,7\}$ with $K=3$ and at $n=7$ with $K=2$. Five instances are generated for each $(n,K)$ combination. The size of the effective Hilbert space for each QWOA-variant is given in \cref{tab:dimensions}. We detail instance generation further in \cref{sec:instance_generation} and provide justification for the choice of $\lambda$ in \cref{sec:penalty_selection}. 

\begin{table*}[t]
\centering
\caption{Effective Hilbert-space dimensions for each simulated $(n,K)$.}
\label{tab:dimensions}
\begin{ruledtabular}
\begin{tabular}{cccccc}
 & & \multicolumn{2}{c}{I-QWOA $|\mathcal H_\text{eff}|$} & \multicolumn{2}{c}{PS-QWOA $|\mathcal H_\text{eff}|$} \\
\cline{3-4}\cline{5-6}
$n$ & $K$
  & \multicolumn{1}{c}{Homogeneous (\cref{eq:hom_solution_count})}
  & \multicolumn{1}{c}{Heterogeneous (\cref{eq:solution_count})}
  & \multicolumn{1}{c}{Homogeneous (\cref{eq:effective_hom_ps_dimension})}
  & \multicolumn{1}{c}{Heterogeneous (\cref{eq:encoding_space_dimension})} \\
\hline
5 & 3 &            480 &          2{,}520 &          4{,}920 &         29{,}160 \\
6 & 3 &          3{,}720 &         20{,}160 &         87{,}840 &        524{,}880 \\
7 & 3 &         32{,}760 &        181{,}440 &      1{,}839{,}600 &     11{,}022{,}480 \\
8 & 3 &        322{,}560 &      1{,}814{,}400 &     44{,}110{,}080 &    264{,}539{,}520 \\
7 & 2 &         20{,}160 &         40{,}320 &             --- &        645{,}120 \\
8 & 2 &        181{,}440 &        362{,}880 &      5{,}160{,}960 &             --- \\
\end{tabular}
\end{ruledtabular}
\end{table*}

\subsubsection{Simulation}
\label{sec:simulation}
Simulations are performed using QuOp\_MPI~\cite{matwiejew_quop_mpi_2021}, which implements an exact simulation of the dynamics in the effective Hilbert space of each QWOA-variant. With this approach, we aim to characterise baseline algorithmic performance on instances larger than those that are tractable with a circuit-based simulation -- which would require the inclusion of ancilla qubits for realisation of the ansatz at a reasonable gate depth. Classically efficient construction of the mixing graphs and computation of the solution costs requires an indexing and unindexing of the algorithm's effective Hilbert space. We detail these indexing schemes in \cref{sec:symmetric_indexing}.

The action of the PS-QWOA permutation mixing unitary in \cref{sec:results} was computed following the first-order Trotter decomposition described in \cref{sec:circuits} with one Trotter repetition per layer. The $\binom{n}{2}$ block-$\mathrm{SWAP}$ terms of $\mathcal G_\pi$ are partitioned into $C$ disjoint matchings via a round-robin edge colouring of $K_n$ (in which the $\binom{n}{2}$ pairs are arranged into $C$ rounds of mutually disjoint pairings), so that the layer walk $e^{-\ii t\,\mathcal G_\pi}$ is implemented as $\prod_{c=0}^{C-1} e^{-\ii t\,\mathcal G_\pi^{(c)}}$. The action of the assignment sub-mixer was computed without Trotterisation, as it has a quantum circuit that computes it exactly at any $t$~\cite{matwiejew_quantum_2024}. Computation of the complete-graph-based I-QWOA mixing unitary utilised FFT-based diagonalisation~\cite{matwiejew_quop_mpi_2021}.

\subsubsection{Optimisation}
\label{sec:optimisation}
For each algorithm and problem instance, simulations were conducted at circuit depths $p=1$ to $8$. The PS-QWOA uses the depth-independent three-parameter linear schedule of \cref{sec:qwoa}. The I-QWOA uses a fully variational parameterisation with $2p$ free parameters $(\gamma_{\ell}, t_{\ell})_{\ell=1}^{p}$, each $\gamma_{\ell}$ normalised by the standard deviation of the penalised cost function~\cite{bennett_quantum_2021}. We do not apply the three-parameter linear schedule to I-QWOA, as its complete-graph mixer violates the structure-based analysis of Ref.~\cite{bennett_non-variational_2024}. The linear schedule is a sufficient rather than optimal parameterisation, so the comparison made here is conservative with respect to the PS-QWOA.

Parameter optimisation is performed with L-BFGS-B (SciPy, $\texttt{maxiter}=10\,000$) with five independent restarts from different initial parameter values at each depth. The I-QWOA restarts are a single warm-start base point plus four independent Gaussian perturbations of that point (standard deviation~$0.1$). At $p\ge 2$ the base is obtained by transferring the depth-$(p{-}1)$ optimum to depth $p$ via the INTERP rule of~\cite{zhou_quantum_2020} (Eq.~B1). Given the depth-$(p{-}1)$ optimum parameter set $(\theta^\star_1,\ldots,\theta^\star_{p-1})$, the depth-$p$ initial guess $(\theta^{(0)}_1,\ldots,\theta^{(0)}_p)$ is the linear interpolation $\theta^{(0)}_i = \tfrac{i-1}{p-1}\,\theta^\star_{i-1} + \tfrac{p-i}{p-1}\,\theta^\star_i$ for $i=1,\ldots,p$, with $\theta^\star_0=\theta^\star_p=0$, applied independently to the $\gamma$ and $t$ parameters. The PS-QWOA reuses the previous-depth optimum as the base for its three-parameter linear schedule (\cref{sec:qwoa}), with Gaussian perturbation of the parameters (standard deviation~$0.1$) after the first repeat. For both algorithms, the best point over the five restarts is retained at each depth and used to warm-start the next depth. 


\subsubsection{Figures of merit}
\label{sec:figures_of_merit}

We evaluate each optimised QWOA state $\ket{\psi(p)}\equiv\ket{\psi(\vec\gamma,\vec t)}$ of \cref{eq:qwoa-ansatz} through a combination of feasibility, solution-quality, and concentration measures. Throughout, $P_x(p)=|\braket{x}{\psi(p)}|^2$ denotes the depth-$p$ measurement probability of basis state $\ket{x}$, and $P(z,p)=\sum_{x\in z}P_x(p)$ for any subset $z\subseteq\Omega$. We use $\mathcal{F}\subseteq\Omega$ for the feasible (capacity-constraint compliant) set, $f^\star=\min_{x\in\mathcal F}f(x)$ for the optimal feasible travel cost, and $\mathcal{Z}^\star=\{x\in\mathcal F:f(x)=f^\star\}$ for the optimal feasible basis states. The effective Hilbert-space dimension $|\mathcal H_{\mathrm{eff}}|$ is the search space of the corresponding algorithm, as summarised in \cref{tab:dimensions}. 

The \emph{feasibility probability} and its initial-state value are
\begin{equation}
\label{eq:p_feas}
P_{\mathrm{feas}}(p) = \sum_{x\in\mathcal F} P_x(p),\qquad P_{\mathrm{feas}}^{(0)} = \sum_{x\in\mathcal F} |\braket{x}{\psi_0}|^2,
\end{equation}
where $P_{\mathrm{feas}}^{(0)}$ is algorithm dependent. The I-QWOA initial state is uniform over the indexed solution space, and the PS-QWOA initial state is uniform over the corresponding effective Hilbert space. Both encode valid routing and assignment structure by construction but may still violate the capacity constraints of \cref{sec:cvrp}, so $P_{\mathrm{feas}}^{(0)}\le 1$ with equality only when every encoded configuration is capacity-feasible. 

Restricting the expected cost to feasible outcomes gives the \emph{feasibility-conditioned approximation ratio}
\begin{equation}
\label{eq:r_feas}
r_{\mathrm{feas}}(p) = \frac{\mathbb{E}[\,f\mid\mathcal F\,]}{f^\star}
 = \frac{\sum_{x\in\mathcal F} f(x)\,P_x(p)}{f^\star\,P_{\mathrm{feas}}(p)}\ge 1.
\end{equation}
We report the reciprocal \emph{feasible quality}
\begin{equation}
\label{eq:quality}
\frac{1}{r_{\mathrm{feas}}(p)}\in(0,1],
\end{equation}
so that a value of one corresponds to deterministic sampling of the feasible optimum and $1/r_{\mathrm{feas}}(p)$ measures routing quality on post-selected feasible outcomes.

For $\alpha\in\{1,\,2,\,5\}$ let $\mathcal T_{\alpha\%}\subseteq\mathcal F$ denote the top $\alpha\%$ of feasible solutions ranked by $f$, with $|\mathcal T_{\alpha\%}|=\max(1,\lceil(\alpha/100)|\mathcal F|\rceil)$. The \emph{top-tier feasible probability} and the corresponding \emph{top-tier amplification} against the uniform baseline are
\begin{equation}
\label{eq:p_alpha}
P_{\alpha\%}(p) = \sum_{x\in\mathcal T_{\alpha\%}} P_x(p),\qquad
A_{\alpha\%}(p) = \frac{|\mathcal H_{\mathrm{eff}}|\,P_{\alpha\%}(p)}{|\mathcal T_{\alpha\%}|}.
\end{equation}
Here $A_{1\%}$ denotes the amplification of the top-$1\%$ feasible tier. $A_{\alpha\%}=1$ corresponds to uniform sampling over $\mathcal H_{\mathrm{eff}}$, and $A_{\alpha\%}>1$ quantifies the factor by which the algorithm concentrates probability on the top-$\alpha\%$ tier beyond that baseline~\cite{bennett_benchmarking_2025,Brassard2002}. We write the optimal-sampling probability as $P_{\mathrm{opt}}(p)=P(\mathcal Z^\star,p)$. The \emph{optimal-feasible amplification} $A_{\mathrm{opt}}(p)$ is defined on the exactly optimal feasible set $\mathcal Z^\star$, replacing the top-$\alpha\%$ tier with $\mathcal Z^\star$.

For each combination of algorithm, instance and depth, we record the per-restart number of objective-function evaluations $N_{\mathrm{fev}}(p)$ (calls to $\langle\mathcal C\rangle_p$ in \cref{eq:objective-function}) by the optimiser. To compare optimiser cost across problem sizes, we use the cumulative count $N_{\mathrm{fev}}^{\mathrm{cum}}=\sum_{p=1}^{8}N_{\mathrm{fev}}(p)$ for the $K=3$ data, first taking the median over restarts at each depth within each instance and then summarising over instances at fixed $(n,K)$. Unless otherwise stated, each figure of merit above is computed per instance and summarised by the median over the five problem instances, and interquartile bands are reported.

\subsection{Results}
\label{sec:results}

\begin{figure*}[t]
\centering
\includegraphics[width=0.85\textwidth]{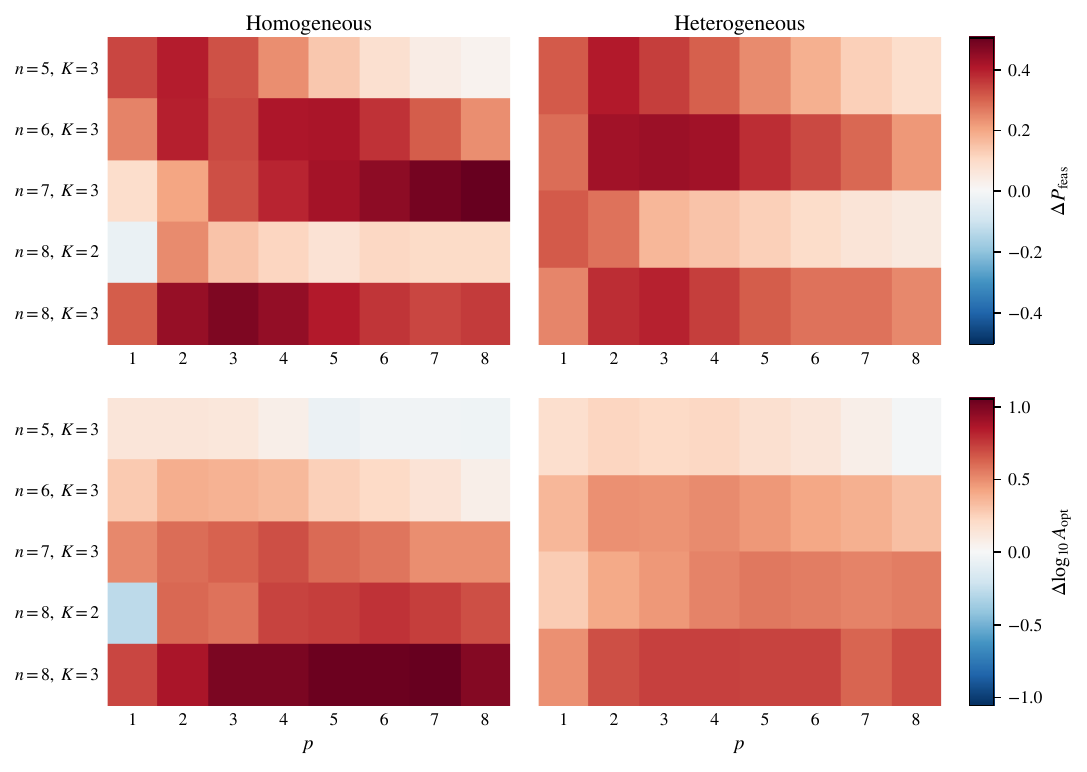}
\caption{Difference between the PS-QWOA and I-QWOA instance medians across the full simulated dataset. Rows are indexed by problem size $(n,K)$ and are ordered by increasing effective Hilbert-space dimension $|\mathcal{H}_{\mathrm{eff}}|$, with the smallest $|\mathcal{H}_{\mathrm{eff}}|$ at the top. Columns are indexed by the circuit depth $p=1,\dots,8$. \textbf{Top row:} Difference in feasibility probability $\Delta P_{\mathrm{feas}}$ for homogeneous CVRP (left) and heterogeneous CVRP (right) instances. \textbf{Bottom row:} Difference in $\log_{10}$ of the optimal-feasible amplification $\Delta\log_{10}A_{\mathrm{opt}}$. Each cell is the difference between the PS-QWOA and I-QWOA medians over the five problem instances. Positive values (red) indicate a PS-QWOA advantage.}
\label{fig:heatmaps}
\end{figure*}

We first summarise how the PS-QWOA compares with the I-QWOA across the entire simulated dataset. \Cref{fig:heatmaps} shows the difference between the feasibility probability $P_{\mathrm{feas}}$ and the optimal feasible amplification $A_{\mathrm{opt}}$ of the two algorithms. Each cell is the instance-median of the PS-QWOA value minus the instance-median of the I-QWOA value, so positive (red) cells indicate a PS-QWOA advantage.

Two patterns are immediately visible. First, the PS-QWOA has an advantage in $P_{\mathrm{feas}}$ for each $(n, K)$ of the heterogeneous and homogeneous problem sets -- aside from the $(n,K)=(8, 2)$ homogeneous instances at $p=1$ where the two algorithms differ by only $-0.032$. Second, $\Delta\log_{10} A_{\mathrm{opt}}$ varies strongly with $|\mathcal{H}_{\mathrm{eff}}|$ (i.e., across rows) but only weakly with depth $p$ (i.e., across columns). On the smallest-$|\mathcal{H}_{\mathrm{eff}}|$ homogeneous row, the advantage of the PS-QWOA is near zero and takes small negative values at the largest depths, whereas on the largest rows the PS-QWOA attains $A_{\mathrm{opt}}$ values that exceed the I-QWOA baseline by factors of up to $\sim\!11$ ($\Delta\log_{10}A_{\mathrm{opt}}=1.05$ at $(n,K,p)=(8,3,7)$) on homogeneous instances, and up to $\sim\!5.3$ ($\Delta\log_{10}A_{\mathrm{opt}}=0.72$ at $(n,K,p)=(7,3,3)$) on heterogeneous instances. The $P_{\mathrm{feas}}$ panel shows a similar, although less pronounced, growth with $|\mathcal{H}_{\mathrm{eff}}|$. On homogeneous instances the median PS-QWOA advantage grows from $\Delta P_{\mathrm{feas}}\approx 0.02$ at the smallest simulated size and $p=8$ to $\approx 0.35$ at the largest $K=3$ size, and the heterogeneous trend is qualitatively the same. Taken together, these two views of the full dataset indicate that the PS-QWOA has a reliable performance advantage over the I-QWOA that strengthens as the effective Hilbert-space dimension grows. 

\begin{figure}[t]
\centering\includegraphics[width=0.8\columnwidth]{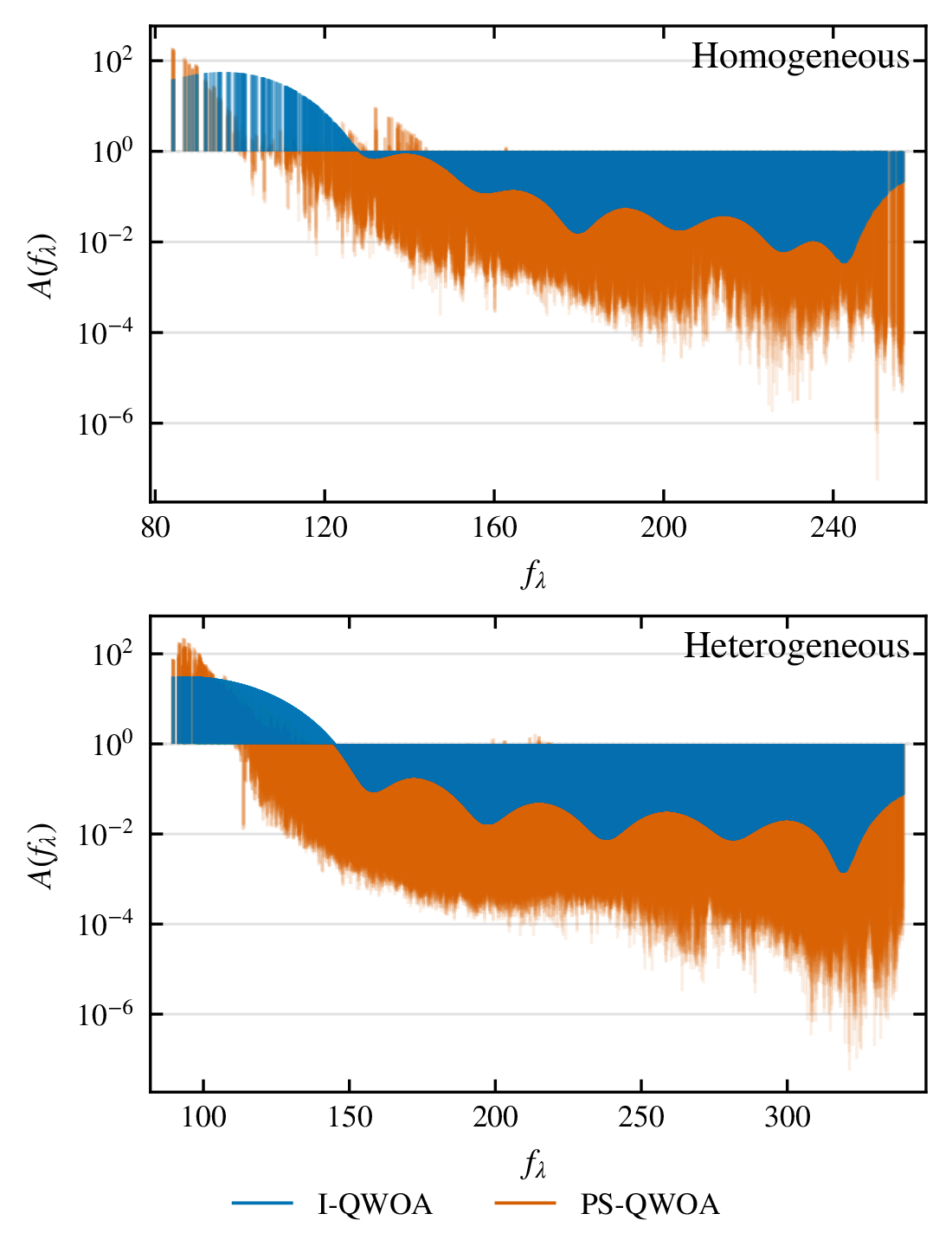}
\caption{Penalised-cost amplification profile $A(f_\lambda)=P(f_\lambda)/P_{\mathrm{unif}}(f_\lambda)$ of optimised PS-QWOA and I-QWOA states for a single problem instance at depth $p=8$. \textbf{Top:} homogeneous CVRP $(n,K)=(7,3)$. \textbf{Bottom:} heterogeneous CVRP $(n,K)=(7,3)$.} 
\label{fig:opt_distributions}
\end{figure}

We next examine the $(n,K)=(7,3)$ problem size in more detail. We begin with one representative instance for each problem variant. \Cref{fig:opt_distributions} shows the penalised-cost amplification profile of prepared I-QWOA and PS-QWOA states for the homogeneous and heterogeneous $(n,K)=(7,3)$ instances that achieved the minimum objective function value at $p=8$. In both cases, the PS-QWOA optimised state concentrates probability more sharply on the lowest-cost feasible region, reaching $A(f_\lambda)>10^2$ near the cost minimum before falling by several orders of magnitude below the uniform baseline across the higher-cost tail. The I-QWOA distribution also amplifies at the lowest costs, but with a less-defined peak that is farther from the minimum and with lower suppression in the higher-cost region. 

\begin{figure*}[!t]
\centering
\includegraphics[width=0.68\textwidth]{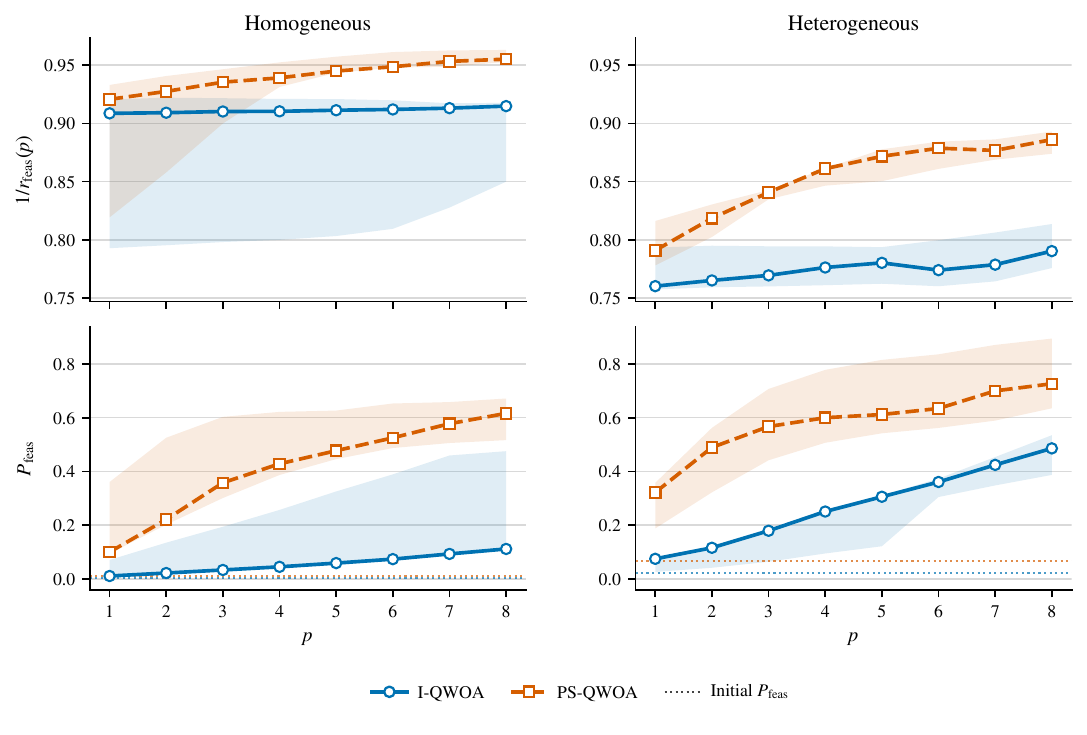}
\caption{Feasibility and feasible-solution quality as a function of circuit depth $p$, for homogeneous CVRP (left, $n=7$, $K=3$) and heterogeneous CVRP (right, $n=7$, $K=3$) instances. Curves show medians over instances, with bands indicating interquartile ranges. \textbf{Top row:} feasible quality $1/r_{\mathrm{feas}}(p)$. \textbf{Bottom row:} feasibility probability $P_{\mathrm{feas}}(p)$, with dotted horizontal lines indicating the initial-state feasibility probability of each algorithm.}
\label{fig:feasibility_quality}
\end{figure*}

Across all $(n,K)=(7,3)$ instances, \cref{fig:feasibility_quality} shows the median feasible quality (\cref{eq:quality}) and median feasibility probability (\cref{eq:p_feas}). The feasible quality improves steadily with $p$ for both algorithms on both problem variants, but the PS-QWOA consistently attains a higher median value. At $p=8$, PS-QWOA reaches $1/r_{\mathrm{feas}}=0.955$ on the homogeneous CVRP $(7,3)$ instances and $0.886$ on the heterogeneous CVRP $(7,3)$ instances, compared with $0.915$ and $0.790$ for I-QWOA respectively. There is a clear gap in feasible quality in favour of the PS-QWOA from $p=1$ to $8$.

The two algorithms prepare different initial states. The I-QWOA, which is uniform over its indexed solution space, gives $P_{\mathrm{feas}}^{(0)}=0.0029$ for the homogeneous CVRP $(n,K)=(7,3)$ and $0.0231$ for the heterogeneous CVRP at the same $n$ and $K$. The PS-QWOA initial state, which is uniform over the product-space encoding Hilbert space, gives $P_{\mathrm{feas}}^{(0)}=0.0110$ and $0.0677$ respectively. Both algorithms increase the feasibility probability as $p$ increases, with the PS-QWOA concentrating more probability mass in this sector at each depth. At $p=8$ the PS-QWOA median reaches $P_{\mathrm{feas}}=0.616$ for the homogeneous CVRP $(7,3)$ and $0.726$ on the heterogeneous CVRP $(7,3)$, compared with $0.112$ and $0.486$ for I-QWOA.

\begin{figure*}[p]
\centering
\includegraphics[width=0.68\textwidth]{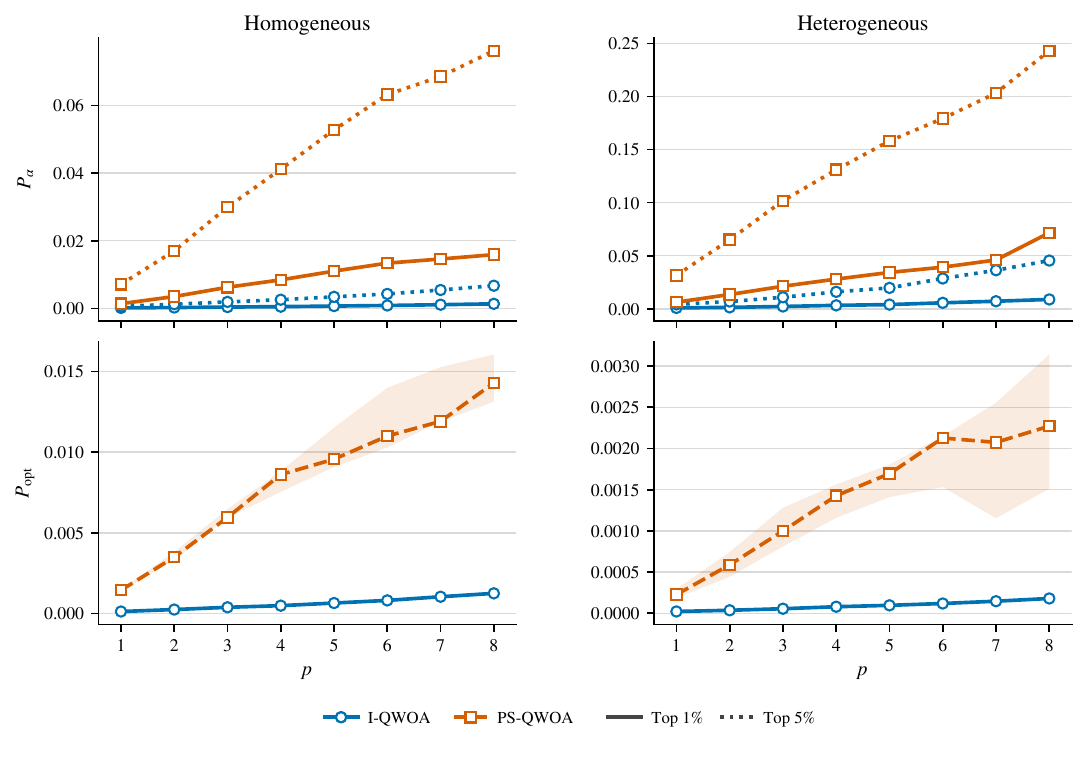}
\caption{High-quality feasible probability and exact optimal-sampling probability as a function of circuit depth $p$, for homogeneous CVRP (left, $n=7$, $K=3$) and heterogeneous CVRP (right, $n=7$, $K=3$) instances. Curves show medians over instances, with bands indicating interquartile ranges. \textbf{Top row:} top-tier feasible probabilities $P_{1\%}(p)$ and $P_{5\%}(p)$. \textbf{Bottom row:} probability $P_{\mathrm{opt}}(p)=P(\mathcal Z^\star,p)$ of sampling an exactly optimal feasible solution.}
\label{fig:concentration}
\end{figure*}

\Cref{fig:concentration} illustrates how much of this difference is attributable to convergence to states in the top $1\%$ and top $5\%$ of the feasible solution space, as measured by $P_{1\%}(p)$ and $P_{5\%}(p)$, as well as the optimal-sampling probability $P_{\mathrm{opt}}(p)$. On the homogeneous CVRP $(7,3)$ instances at $p=8$, the PS-QWOA median concentrates $0.0159$ and $0.0761$ in the top $1\%$ and top $5\%$ feasible tiers, compared with $0.00134$ and $0.0067$ for I-QWOA. On the heterogeneous CVRP $(7,3)$ instances, the corresponding PS-QWOA values are $0.0716$ and $0.243$, compared with $0.0089$ and $0.045$ for I-QWOA. A qualitatively similar separation persists for the exact optimum. At $p=8$, PS-QWOA yields $P_{\mathrm{opt}}=1.43\times 10^{-2}$ on the homogeneous CVRP $(7,3)$ and $2.27\times10^{-3}$ on the heterogeneous $(7,3)$, compared with $1.25\times10^{-3}$ and $1.79\times10^{-4}$ for I-QWOA. Altogether, we observe that the product-space walk not only shifts probability into a broad, high-quality feasible tier but also substantially increases the probability of sampling an exactly optimal route.

\begin{figure*}[!t]
\centering
\includegraphics[width=0.80\textwidth]{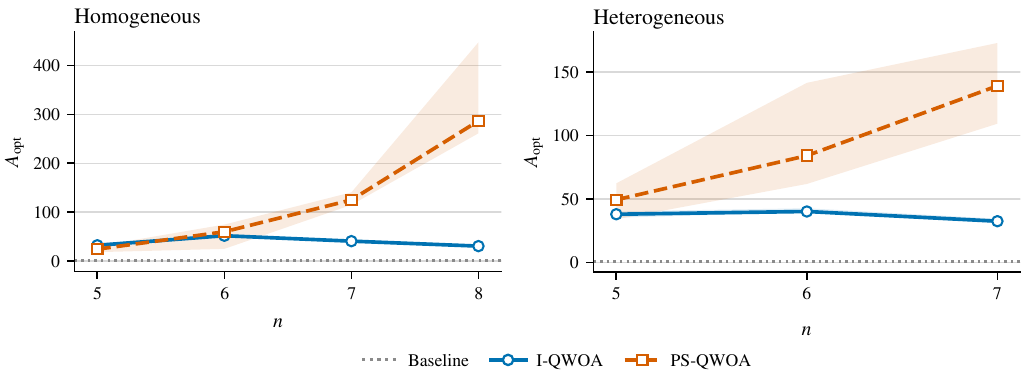}
\caption{Optimal-feasible amplification $A_{\mathrm{opt}}$ at fixed depth $p=8$ as a function of the number of customers $n$ for $K=3$ instances. The dotted line marks the uniform-sampling baseline $A_{\mathrm{opt}}=1$.}
\label{fig:opt_amp_scaling}
\end{figure*}

Returning to the full $K=3$ size series, \cref{fig:opt_amp_scaling} reports the amplification of the optimal set $A_{\mathrm{opt}}$ at fixed depth $p=8$ as a function of $n$. In the homogeneous case, the PS-QWOA median increases from $24$ at $n=5$ to $287$ at $n=8$, whereas I-QWOA remains between $31$ and $52$, decreasing from $n=6$ to $8$. In the heterogeneous case, the PS-QWOA median increases from $49$ at $n=5$ to $139$ at $n=7$, while I-QWOA remains near $30$--$40$. In each case, we observe that the gap in the amplification of the optimal set widens in favour of the PS-QWOA as the problem size increases, even though the PS-QWOA has the larger effective Hilbert space out of the two ans\"atze.


\begin{figure*}[!t]
\centering
\includegraphics[width=0.80\textwidth]{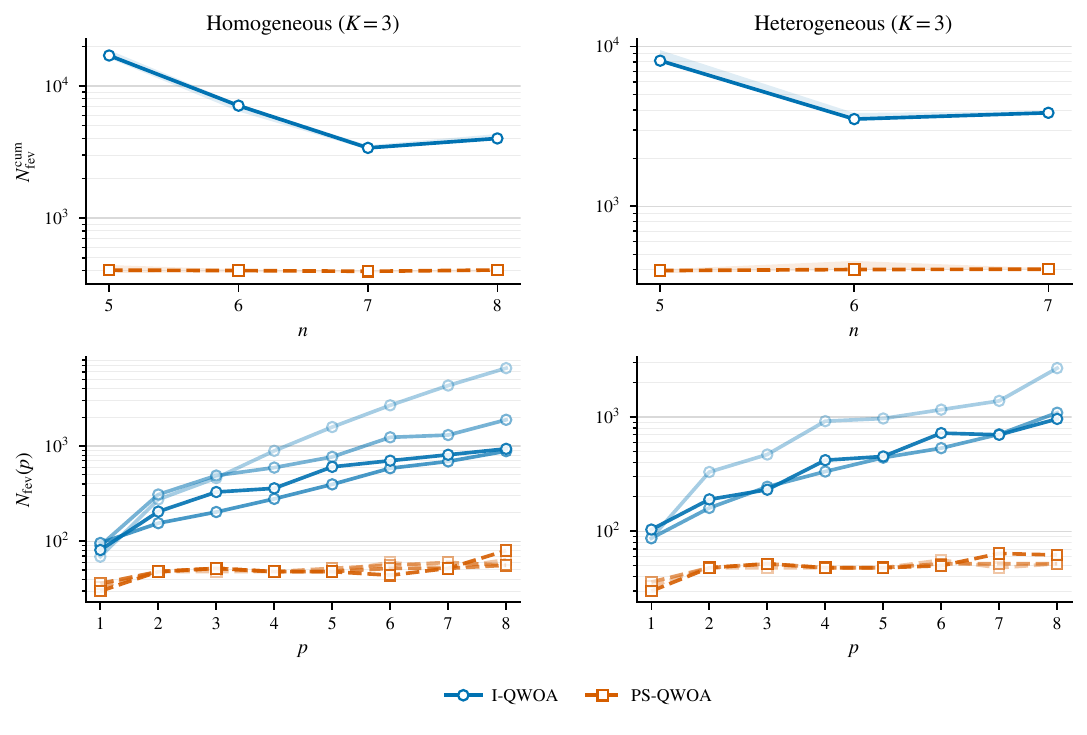}
\caption{Classical optimiser cost for $K=3$ homogeneous CVRP (left) and heterogeneous CVRP (right) instances. \textbf{Top row:} $N_{\mathrm{fev}}^{\mathrm{cum}}=\sum_{p=1}^{8}N_{\mathrm{fev}}(p)$, the cumulative number of L-BFGS-B objective-function evaluations, as a function of the number of customers $n$. \textbf{Bottom row:} depth-wise objective-function evaluations $N_{\mathrm{fev}}(p)$ as a function of circuit depth $p$ for each available $n$ value, with lighter curves denoting smaller $n$ and darker curves denoting larger $n$. The objective function is the ansatz expectation value in \cref{eq:objective-function}.}
\label{fig:optimizer_cost}
\end{figure*}

Finally, \cref{fig:optimizer_cost} shows the cumulative objective-function evaluation count $N_{\mathrm{fev}}^{\mathrm{cum}}=\sum_{p=1}^{8}N_{\mathrm{fev}}(p)$ as a function of $n$ and the corresponding depth-wise counts $N_{\mathrm{fev}}(p)$ for the $K=3$ problem instances. The PS-QWOA cumulative cost is nearly size-independent, remaining close to $4\times10^{2}$ objective function evaluations for both problem variants. The I-QWOA cumulative cost is much larger, ranging from $3.4\times10^{3}$ to $1.7\times10^{4}$ on the homogeneous $K=3$ instances and from $3.5\times10^{3}$ to $8.1\times10^{3}$ on the heterogeneous $K=3$ instances. Notably, the I-QWOA cumulative count is non-monotone in $n$ over the simulated range, with the largest values occurring at the smallest size $n=5$ ($1.7\times10^{4}$ on homogeneous and $8.1\times10^{3}$ on heterogeneous instances) and decreasing to $\sim\!4\times10^{3}$ at $n\ge 7$. This is consistent with the depth-wise profiles in the lower panels, where the I-QWOA per-depth count is highest at $n=5$ and decreases as $n$ grows, reflecting the optimiser terminating sooner from the warm-started point as the landscape becomes flatter near the previous-depth optimum at larger sizes. The lower panels also show that the gap between the two algorithms is driven by the steep increase in the I-QWOA evaluation count with depth, whereas the PS-QWOA count remains nearly flat. At the highest simulated $K=3$ sizes, the PS-QWOA cumulative cost is nearly an order of magnitude below the I-QWOA value at the same $n$ ($\sim\!8.6\times$ on homogeneous $(7,3)$ and $\sim\!9.5\times$ on heterogeneous $(7,3)$).


\section{Discussion}
\label{sec:discussion}

The PS-QWOA reduces the per-layer gate complexity of the earlier indexed QWOA formulation (I-QWOA)~\cite{bennett_quantum_2021} from $\mathcal{O}(n^3\log n)$ to $\mathcal{O}(n^2\log n)$, a factor-of-$n$ improvement at comparable qubit count. To express qubit cost as a single-parameter worst-case scaling, one may take $K=n$, the largest meaningful fleet size, since there can be at most $n$ nonempty routes. In that regime, both formulations use $\Theta(n\log n)$ data qubits, and the canonical solution-space dimensions in the heterogeneous and homogeneous cases (\cref{eq:solution_count,eq:hom_solution_count}) satisfy $\log|\Omega|\approxeq\Theta(n\log n)$ by Stirling's approximation~\cite{petkovsek2007lah}. The computational complexity of the I-QWOA is dominated by its indexing unitary $U_\#$ (see \Cref{sec:qwoa}), which requires $\mathcal{O}(n^3\log n)$ two-qubit gates when implemented with standard ripple-carry arithmetic primitives~\cite{Cuccaro2004}, with the same scaling applying to the heterogeneous case as the indexing algorithm has the same asymptotic complexity (\cref{sec:symmetric_indexing}). The product-space encoding of the PS-QWOA sidesteps the need for the indexing unitary altogether, as it naturally encodes the vehicle assignment constraint and accommodates heterogeneous fleets without modification.

The approach taken here also contrasts with the QAOA formulation of Fitzek et al.~\cite{Fitzek2024}, which addresses a closely related heterogeneous-fleet routing variant and encodes it in $\mathcal{O}(n^2K)$ qubits using route-position binary variables, with routing and capacity constraints enforced by penalty terms and a transverse-field mixer over the full binary state space. By our determination, assuming a standard gate-model compilation in which the diagonal cost Hamiltonian is expanded into one- and two-body $Z$ terms and each quadratic term is implemented using only constant-size two-qubit entangling gadgets, the resulting per-layer two-qubit gate complexity is $\mathcal{O}(Kn^4 + n^3K^2)$. Furthermore, at depth $p=5$, this QAOA formulation is unable to distinguish between feasible solutions in the two larger instances, a failure mode that the PS-QWOA avoids.

The numerical results in \cref{sec:results} show substantially improved optimisation performance. In brief, PS-QWOA attains higher feasibility probabilities, higher feasible-solution quality, and higher concentration on the top-quality and optimal feasible tiers than I-QWOA across both problem variants, with the strongest homogeneous gains appearing from $p\ge 2$ onward and heterogeneous gains evident at all simulated depths (\cref{fig:feasibility_quality,fig:concentration,fig:opt_amp_scaling}). Over the same range, we find that the PS-QWOA amplification of feasible and globally optimal solutions increases with the dimension of its effective Hilbert space, whereas the complete-graph I-QWOA degrades. 

All of this is observed while parameterising the PS-QWOA via the three-parameter linear schedule of~\cite{bennett_non-variational_2024} (\cref{sec:qwoa}), which keeps the objective-function evaluation count approximately constant as depth and problem size vary. In contrast, the optimisation over the full $2p$ parameters in the original I-QWOA is substantially more expensive over all problem instances. The ability of the linear schedule to produce sufficiently optimal $\gamma$ and $t$ values indicates that the PS-QWOA is successfully exploiting global structure in the problem search space~\cite{bennett_non-variational_2024}, and extends this "non-variational" parameterisation scheme to composite mixing unitaries that act on distinct combinatorial structures.


Limitations do apply to the presented results. The problem sizes considered ($n\le 8$) are small compared to the number of vehicles and customers in production CVRP instances. So, while the observed scaling trends are encouraging, extrapolation to practically relevant instance sizes warrants caution. In addition, the simulations use exact state-vector evolution for the encoded walk generators and cost operators, which does not account for hardware noise. Nevertheless, the performance advantage of the PS-QWOA over I-QWOA is both consistent and pronounced, placing it as the favourable candidate for quantum walk-based optimisation of the CVRP.

\section{Conclusion}
\label{sec:conclusion}

We have presented a product-space quantum walk-based optimisation algorithm (PS-QWOA) for the capacitated vehicle routing problem. The method encodes solutions as a product of the spaces of customer orderings and vehicle assignments, using structured mixing unitaries derived from the combinatorial neighbourhood of each space. This encoding accommodates heterogeneous vehicle fleets with vehicle-specific capacities and costs, extending the applicability of the QWOA framework beyond the homogeneous case treated in prior work~\cite{bennett_quantum_2021}, while also reducing the overall gate complexity.

Numerical simulations on CVRP instances with up to $n = 8$ customers and $K = 3$ vehicles show that PS-QWOA, equipped with a depth-independent three-parameter linear schedule, outperforms the I-QWOA at all depths on heterogeneous instances and from $p \ge 2$ onward on the homogeneous instances. The PS-QWOA consistently delivers a substantially higher optimal-sampling probability and concentrates markedly more probability on the lowest-cost solutions that satisfy the capacity constraints, with lower classical optimisation overhead. Across the simulated dataset, the PS-QWOA advantage broadens with problem size, while its optimiser cost remains nearly size-independent over the range simulated.

Future work will explore circuit-based simulations incorporating realistic noise models and extend the tensor-product encoding to other scheduling and resource-allocation problems with analogous product-space structures.

\section*{Acknowledgements}

This project was funded through the Advanced Strategic Capabilities Accelerator (ASCA). Resources were provided by the Pawsey Supercomputing Research Centre's Setonix Supercomputer (https://doi.org/10.48569/18sb-8s43), which received funding from the Australian Government and the Government of Western Australia. Additional support came from Pawsey's Quantum Supercomputing Innovation Hub, made possible by a grant from the Australian Government through the National Collaborative Research Infrastructure Strategy (NCRIS).

\bibliographystyle{apsrev4-2}
\bibliography{bibliography}

\begin{appendix}
\crefalias{section}{appendix}

\section{Reduced basis and walk generators in the symmetric subspace}
\label{sec:orbit_derivation}

\subsection{Relabelling-symmetric subspace}

The action of the symmetric group $S_K$ on the assignment register is $g\ket{\pi,a}=\ket{\pi,g(a)}$ with $(g(a))_j=g(a_j)$ for $g\in S_K$, and the \emph{relabelling-symmetric subspace} is
\begin{equation}
\label{eq:sym_subspace_def}
  \mathcal H_a^{\mathrm{sym}}
     = \bigl\{\ket{\psi}\in\mathcal H_a : g\ket{\psi}=\ket{\psi}\ \forall g\in S_K\bigr\}.
\end{equation}
Both walk generators $\mathcal G_\pi$ and $\mathcal G_a$ commute with this action because each is a sum of edge terms $\ket{\pi', a'}\bra{\pi,a}+\ket{\pi,a}\bra{\pi', a'}$ whose adjacency structure is invariant under global vehicle relabelling. If $\mathcal G$ commutes with $g$, so does its matrix exponential, and for any $\ket{\psi}\in\mathcal H_\pi \otimes \mathcal H_a^{\mathrm{sym}}$,
\begin{equation}
\label{eq:sym_preservation}
g\,e^{-\ii t\mathcal G}\ket{\psi}
   = e^{-\ii t\mathcal G}\,g\ket{\psi}
   = e^{-\ii t\mathcal G}\ket{\psi},
\end{equation}
so the mixing unitaries preserve $\mathcal H_\pi \otimes \mathcal H_a^{\mathrm{sym}}$. Consequently, if the cost operator is invariant under vehicle relabelling, the dynamics of the PS-QWOA remain in $\mathcal H_\pi \otimes \mathcal H_a^{\mathrm{sym}}$.

\subsection{Canonical representatives and reduced basis}

Given identical vehicles, an equivalence class of vehicle assignments can be uniquely represented by a \emph{canonical restricted growth string} (RGS) $s=(s_1,\dots,s_n)$ satisfying $s_1=0$, $s_i\le 1+\max(s_1,\dots,s_{i-1})$, and $\max_i s_i<K$~\cite{Knuth_TAOCP,NijenhuisWilf1978}. Positions sharing the same RGS value are assigned to the same vehicle, and the number of distinct canonical RGS is $M(n,K)$. The homogeneous PS-QWOA effective Hilbert space is
\begin{equation}
\label{eq:effective_hom_ps_subspace}
  \mathcal H_\pi \otimes \mathcal H_a^{\mathrm{sym}}
      \;\simeq\;
      \operatorname{span}\!\left\{\ket{\pi,s} : \pi\in S_n,\,s\in\mathcal{R}_{n,K}\right\},
\end{equation}
where $\mathcal{R}_{n,K}$ denotes the set of canonical RGS of length $n$ with at most $K$ symbols. Its dimension $n!\,M(n,K)$ exceeds the number of unique homogeneous solutions $|\Omega_{\mathrm{hom}}|$ (\cref{eq:hom_solution_count}) because different basis states represent distinct visit-order interleavings that produce the same ordered routes. 

For example, at $n=3$ customers and $K=2$ vehicles, $(\pi=(1,2,3),\,a=(1,1,2))$ and $(\pi=(1,2,3),\,a=(2,2,1))$ both encode the solution $\{(0,1,2,0),\,(0,3,0)\}$, with both corresponding to the canonical RGS $s=(0,0,1)$. There are $n!\,K^n=48$ labelled states but only $n!\,M(n,K)=24$ relabelling-symmetric basis states, corresponding to $M(3,2)=4$ canonical assignment classes. These 24 basis states encode $|\Omega_{\mathrm{hom}}|=\sum_k \operatorname{Lah}(3,k)=12$ unique solutions.

The probability of measuring a permutation $\pi$ and a canonical assignment $s$ is
\begin{equation}
  \mathrm{Prob}(\pi,s)=\sum_{a\in[s]}\bigl|\langle \pi,a\mid \psi(\vec\gamma,\vec t)\rangle\bigr|^2,
\end{equation}
where $[s]$ is the orbit of $s$. By the orbit--stabiliser theorem,\footnote{For a group $G$ acting on a set $X$, the orbit--stabiliser theorem states $|G|=|Gx|\cdot|\mathrm{Stab}(x)|$ for any $x\in X$~\cite{cameron1999permutation}. Here $G=S_K$ and $x=s$, and the stabiliser consists of all permutations of the $K-m$ unused vehicle labels, giving $|\mathrm{Stab}(s)|=(K-m)!$.} $|\mathrm{Stab}(s)|=(K-m)!$ where $m=|\{s_1,\dots,s_n\}|$ is the number of distinct symbols in $s$, so the orbit size is the falling factorial
\begin{equation}
\label{eq:orbit_size}
  |[s]| = \frac{K!}{(K-m)!} = K^{\underline{m}}.
\end{equation}
Denoting the normalised equal-weight superposition over the orbit $[s]$ by
\begin{equation}
\label{eq:orbit_state}
  \ket{s} \equiv |[s]|^{-1/2}\sum_{a\in[s]}\ket{a},
\end{equation}
so that $\ket{\pi,s}=\ket{\pi}\otimes\ket{s}$. Since the full initial state $\ket{\psi_0}$ is $S_K$-invariant, grouping the assignment sum by orbit yields its reduced-basis form,
\begin{equation}
\label{eq:initial_state_sym}
  \ket{\psi_0}
  = \frac{1}{\sqrt{n!\,K^n}}
    \sum_{\pi\in S_n}\sum_{s\in\mathcal{R}_{n,K}}
    \sqrt{|[s]|}\,\ket{\pi,s}.
\end{equation}

\subsection{Restricted walk generators}

Since $\mathcal G_\pi$ acts only on the permutation register, its restriction to each canonical RGS $s\in\mathcal{R}_{n,K}$ is a disjoint copy of the symmetric Cayley graph of $S_n$.

The matrix elements of $\mathcal G_a$ in the orbit basis $\{\ket{s}\}$ can be expressed in terms of a transition count $T(s\to s')$ that tallies the single-position reassignments of any representative $a\in[s]$ that land in the orbit $[s']$:
\begin{equation}
\label{eq:transition_count}
\begin{aligned}
 T(s\to s') = \bigl|\bigl\{(p,v) : {}&1\le p\le n,\; v\in\{1,\dots,K\},\\
   &v\ne a_p,\; \mathrm{can}(a\oplus_p v) = s'\bigr\}\bigr|.
\end{aligned}
\end{equation}
where $a\oplus_p v$ denotes $a$ with position $p$ changed to $v$ and $\mathrm{can}(\cdot)$ returns the canonical RGS. Because $S_K$ acts transitively on each orbit, $T(s\to s')$ is independent of the representative $a$ chosen. Summing over all $|[s]|$ orbit elements gives $|[s]|\,T(s\to s')$ total transitions, so
\begin{equation}
\label{eq:orbit_matrix_element}
\bra{s'}\mathcal{G}_a\ket{s}
  = \frac{T(s\to s')}{n(K{-}1)}\sqrt{\frac{|[s]|}{|[s']|}}.
\end{equation}
When $K=2$, all orbits have size $K!=2$ and no self-loops arise, so every vertex has degree $n$ in the orbit graph with uniform weight $1/n$ and unit total coupling. For $K\ge 3$, orbit sizes vary and self-loops can appear (when a position is reassigned to a label already present in an equivalent assignment), so the total coupling per vertex depends on the orbit. The reduced assignment graph and the resulting Cartesian-product structure are illustrated for $n=3$ and $K=2$ in \cref{fig:symmetric_subspace_graphs_n3k2}. The permutation register graph is the transposition graph on $S_3$ shown in \cref{fig:walk_generator_graphs_n3k2}.

\input{figures/symmetric_subspace_graphs}

\section{Cost Oracle Circuit}
\label{sec:cost_oracle_details}

Here we detail the implementation of the phase-separation unitary $U_{\mathcal{C}}(\gamma)$ on a gate-based quantum processor. All classical data lookups use QROAM~\cite{Babbush2018encoding,Berry2019qubitization}, which loads an entry from a table of $N$ values using $\mathcal{O}(\sqrt{N})$ Toffoli gates and $\mathcal{O}(\sqrt{N}\cdot b)$ ancilla qubits, where $b$ is the arithmetic-register width.

In the $(\pi,a)$ encoding, the travel cost is a sum over $n+1$ edge contributions obtained by scanning the positions in the order specified by $\pi$ and grouping consecutive positions with the same vehicle label. Each edge requires a distance lookup from the $(n{+}1)\times(n{+}1)$ matrix $c_{ij}$ at $\mathcal{O}(n)$ Toffoli gates per lookup~\cite{Babbush2018encoding}, an equality test on adjacent assignment labels at $\mathcal{O}(\log K)$ gates~\cite{Barenco1995}, and a controlled phase rotation conditioned on the $b$-bit result register at $\mathcal{O}(b)$ gates. For the heterogeneous CVRP, each edge additionally requires a vehicle modifier lookup from a $K$-entry table at $\mathcal{O}(\sqrt{K})$ gates and a multiplication $\alpha_k\times c_{ij}$ at $\mathcal{O}(b^2)$ gates~\cite{Vedral1996}. Over all $n+1$ edges, the travel cost requires $\mathcal{O}(n^2 + n\sqrt{K} + nb^2)$ gates and $\mathcal{O}(nb)$ ancillas for the heterogeneous CVRP, reducing to $\mathcal{O}(n^2)$ gates for the homogeneous CVRP.

The capacity penalty accumulates for each vehicle. At each of the $n$ positions, the demand $d_{\pi(j)}$ is loaded from an $n$-entry demand table at $\mathcal{O}(\sqrt{n})$ gates per lookup and added to one of $K$ accumulator registers of $b$ bits each, conditioned on the assignment label $a_j$, using $\mathcal{O}(Kb)$ controlled additions~\cite{Cuccaro2004}. After accumulation, each vehicle's total demand $D_k$ is compared against its capacity threshold, and any excess is applied as a penalty phase at $\mathcal{O}(b)$ gates per vehicle~\cite{Vedral1996}. For the homogeneous CVRP, the threshold $Q$ is a single classical constant hardcoded into all $K$ comparators. For the heterogeneous CVRP, each threshold $Q_k$ is vehicle-dependent and loaded sequentially at $\mathcal{O}(\sqrt{K})$ gates per threshold, adding $\mathcal{O}(K\sqrt{K})$ gates in total. The capacity penalty thus requires $\mathcal{O}(n\sqrt{n} + nKb)$ gates and $\mathcal{O}(Kb)$ ancillas for the homogeneous CVRP, with an additional $\mathcal{O}(K\sqrt{K})$ gates for the heterogeneous CVRP.

\section{Classical indexing of the state space}
\label{sec:symmetric_indexing}

\begin{figure*}
\centering
\includegraphics[width=0.85\textwidth]{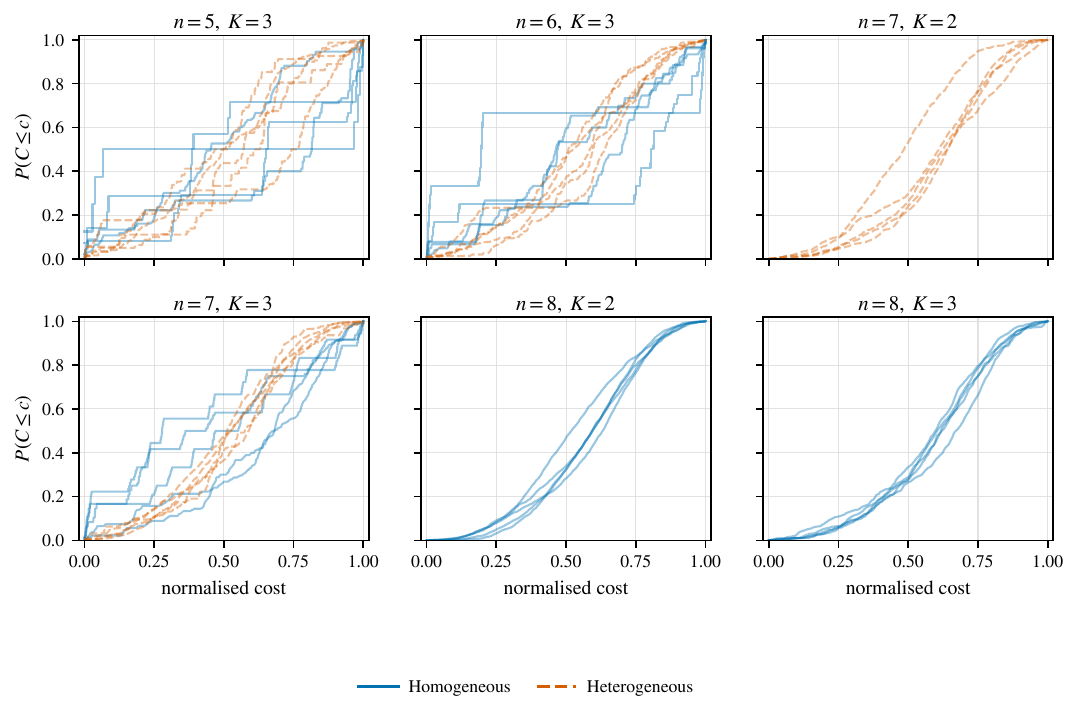}
\caption{Cumulative distributions of the normalised objective over the feasible solutions of each generated instance. Each panel corresponds to a problem size $(n,K)$ simulated in \cref{sec:numerical}. Blue solid and orange dashed curves show homogeneous CVRP and heterogeneous CVRP instances, respectively.}
\label{fig:instance_distributions}
\end{figure*}

Memory-efficient classical simulation of the QWOA requires a bijection between each index of the effective Hilbert space and its corresponding state in the encoding basis. Here we describe the indexing schemes used in this work. These address the full product space $S_n\times\{1,\dots,K\}^n$ (used by PS-QWOA for the heterogeneous CVRP), the reduced relabelling-symmetric subspace $S_n\times\mathcal{R}_{n,K}$ (used by PS-QWOA for the homogeneous CVRP), the canonical heterogeneous solution space $S_n\times\mathrm{Comp}(n,K)$ (used by I-QWOA for the heterogeneous CVRP), and the bijective homogeneous canonical solution space of dimension $|\Omega_{\mathrm{hom}}|=\sum_k \operatorname{Lah}(n,k)$ (\cref{eq:hom_solution_count}), used by I-QWOA for the homogeneous CVRP~\cite{bennett_quantum_2021}. Each scheme is invertible, enabling translation between the indexed and unindexed state representations as required. 

The PS-QWOA and I-QWOA for the heterogeneous CVRP share the same permutation indexing subroutine. A permutation $\pi\in S_n$ is converted to an integer $q\in\{0,\dots,n!-1\}$ via the factoradic (Lehmer code) representation~\cite{Knuth_TAOCP}. Write
\begin{equation}
q = \sum_{j=1}^{n-1} \ell_j\,(n{-}j)!, \qquad 0\le \ell_j\le n{-}j,
\end{equation}
and build $\pi$ by selecting, at each step $j$, the $\ell_j$-th remaining element from a list initially containing $\{1,\dots,n\}$. The inverse (ranking) recovers $q$ from $\pi$ by counting inversions.

In the full product space $S_n\times\{1,\dots,K\}^n$, each state $(\pi,a)$ is indexed by
\begin{equation}
\begin{aligned}
I &= q\,K^n + \rho,\\
q &\in\{0,\dots,n!-1\},\quad \rho\in\{0,\dots,K^n-1\}.
\end{aligned}
\end{equation}
where $q$ is the rank of $\pi$ and $\rho$ is the mixed-radix representation of the assignment $a$, with $\rho=\sum_{j=1}^{n}(a_j-1)\,K^{n-j}$~\cite{Knuth_TAOCP}.

For heterogeneous I-QWOA, the effective Hilbert space is the canonical solution space $S_n\times\mathrm{Comp}(n,K)$ of dimension $n!\binom{n+K-1}{K-1}$, where $\mathrm{Comp}(n,K)$ denotes the set of weak compositions of $n$ into $K$ non-negative parts. A weak composition $u=(u_1,\dots,u_K)$ with $\sum_k u_k=n$ encodes the route lengths, so that vehicle $k$ serves the $u_k$ consecutive customers at positions $\sum_{j<k}u_j+1$ through $\sum_{j\le k}u_j$ in the permutation $\pi$~\cite{Stanley2011EC1}. Each state $(\pi,u)$ is indexed by
\begin{equation}
\begin{aligned}
I &= q\,\binom{n+K{-}1}{K{-}1} + \rho,\\
q &\in\{0,\dots,n!{-}1\},\quad
\rho\in\Bigl\{0,\dots,\binom{n+K{-}1}{K{-}1}{-}1\Bigr\}.
\end{aligned}
\end{equation}
where $q$ is the rank of $\pi$ and $\rho$ is the lexicographic rank of $u$ among all weak compositions~\cite{Knuth_TAOCP}. The rank $\rho$ is computed left-to-right. At position $k$, with $K{-}k$ remaining parts and residual sum $s_k = n - \sum_{j<k}u_j$, the number of compositions preceding those with $u_k=v$ is $\binom{s_k - v + K - k - 1}{K - k - 1}$, giving
\begin{equation}
\rho = \sum_{k=1}^{K-1} \sum_{v=0}^{u_k - 1} \binom{s_k - v + K - k - 1}{K - k - 1}.
\end{equation}

For homogeneous PS-QWOA, the effective subspace is $S_n\times\mathcal{R}_{n,K}$ with dimension $n!\,M(n,K)$, where $\mathcal{R}_{n,K}$ is the set of canonical RGS defined in \cref{sec:cvrp_reduction}. Each state $(\pi,s)$ is indexed by
\begin{equation}
\begin{aligned}
I &= q\,M(n,K) + \rho,\\
q &\in\{0,\dots,n!-1\},\quad \rho\in\{0,\dots,M(n,K)-1\}.
\end{aligned}
\end{equation}
where $q$ is the rank of $\pi$ as before and $\rho$ is the rank of $s$ among the canonical RGS. The permutation component uses the same factoradic scheme described above. The RGS rank $\rho$ is computed by scanning the string $s=(s_1,\dots,s_n)$ left-to-right against a precomputed table of Stirling numbers~\cite{Knuth_TAOCP,NijenhuisWilf1978}. The number of valid RGS prefixes of length $\ell$ using exactly $m$ distinct labels is the Stirling number of the second kind $S(\ell,m)$, which satisfies the recurrence~\cite{Stanley2011EC1}
\begin{equation}
S(\ell,m)=S(\ell{-}1,m{-}1)+m\,S(\ell{-}1,m),
\end{equation}
with $S(0,0)=1$ and $S(\ell,0)=0$ for $\ell>0$. Altogether, the per-state indexing and unindexing operations require $\mathcal{O}(n^2)$ arithmetic steps for the full product space, $\mathcal{O}(n^2+K)$ for the canonical heterogeneous space, and $\mathcal{O}(n^2+nK)$ for the homogeneous PS-QWOA effective subspace, with an additional one-time $\mathcal{O}(nK)$ precomputation of the Stirling table~\cite{Knuth_TAOCP,NijenhuisWilf1978}.

Finally, for homogeneous I-QWOA, solutions are ordered set partitions of $\{1,\dots,n\}$ into at most $K$ nonempty, totally ordered subsets, identified up to permutation of the vehicle labels. The indexing exploits the Lah-number recurrence~\cite{petkovsek2007lah}
\begin{equation}
\label{eq:lah_recurrence}
\operatorname{Lah}(n,k)=\operatorname{Lah}(n{-}1,k{-}1) + (n{+}k{-}1)\,\operatorname{Lah}(n{-}1,k),
\end{equation}
with $\operatorname{Lah}(n,n)=1$ and $\operatorname{Lah}(n,k)=0$ for $k>n$ or $k=0$, which decomposes the $\operatorname{Lah}(n,k)$ solutions with exactly $k$ routes according to whether the largest-labelled customer $n$ occupies a singleton route or is inserted into one of $n+k-1$ positions among the remaining $k$ routes. By recursively peeling off the largest element and accumulating at each step the count of solutions that precede the current one, both the indexing and unindexing operations can be performed in $\mathcal{O}(n^2)$ arithmetic steps. Full details for the homogeneous I-QWOA indexing algorithm are given in Ref.~\cite{bennett_quantum_2021}.

\section{Problem instances}
\label{sec:instance_generation}

Customer coordinates are drawn from three Gaussian clusters (centres uniform in $[-8,8]^{2}$, standard deviation $1.5$, with one outlier relocated $12$--$16$ units) for homogeneous instances, and uniformly from $[-10,10]^{2}$ for heterogeneous instances. The pairwise cost matrix is constructed from Euclidean distances with a $1\%$ antisymmetric random perturbation, so that a route and its reverse incur distinct costs. Each customer receives an integer demand drawn uniformly from $\{1,\dots,5\}$. Homogeneous instances assign all vehicles a shared capacity $Q = \max(6,\lceil 3n/K \rceil + 1)$, while heterogeneous instances assign each vehicle an individual capacity $\lceil\text{total demand}/K\rceil \pm 2$ and a cost modifier $\alpha_{k}\sim\mathrm{Uniform}[0.8,\,1.5]$. In both cases, demands are adjusted as needed to ensure that feasible solutions exist. \Cref{fig:instance_distributions} shows the resulting cumulative cost distributions. For each instance we normalise cost to $[0,1]$ by $(C-C_{\min})/(C_{\max}-C_{\min})$ and plot $P(C\le c)$ under the uniform distribution over feasible solutions. Homogeneous and heterogeneous instances with the same $(n, K)$ are overlaid.

\section{Penalty-weight selection}
\label{sec:penalty_selection}

We consider five candidate rules for the penalty weight $\lambda$ in the penalised objective $f_\lambda$ of \cref{eq:penalised-cost}. All share a common prefactor
\begin{equation}
\kappa = s \, \max_{k}\alpha_{k},
\label{eq:penalty_prefactor}
\end{equation}
where $s>0$ is an optional user multiplier (set to $s=1$ throughout) and $\max_{k}\alpha_{k}$ reduces to unity for homogeneous fleets. In the rules below, the mean, median and maximum are taken over the off-diagonal costs $c_{ij}$, with indices ranging over the depot and customer nodes $\{0,\dots,n\}$. With $d_i$ denoting the demand of customer $i$, the five rules are
\begin{align}
\lambda_{\text{med}} &= \kappa\,\operatorname*{median}_{i\ne j} c_{ij}, \label{eq:lambda_med}\\
\lambda_{\text{mean}}&= \kappa\,\operatorname*{mean}_{i\ne j} c_{ij}, \label{eq:lambda_mean}\\
\lambda_{\max}       &= \kappa\,\max_{i\ne j} c_{ij}, \label{eq:lambda_max}\\
\lambda_{\text{route}}&= \kappa\,\frac{(n+1)\operatorname*{median}_{i\ne j} c_{ij}}{\tfrac{1}{n}\sum_{i=1}^{n} d_{i}}, \label{eq:lambda_route}\\
\lambda_{\text{sep}}  &= \kappa\,(n+1)K\,\max_{i\ne j} c_{ij} + 1, \label{eq:lambda_sep}
\end{align}
Here $\lambda_{\text{sep}}$ is a strict big-$M$ upper bound on feasible travel cost, so the feasible and infeasible cost spectra are disjoint; the remaining four are soft penalties whose scale tracks a worst-case \eqref{eq:lambda_max}, per-route \eqref{eq:lambda_route}, or typical \eqref{eq:lambda_mean},\eqref{eq:lambda_med} edge cost, and keep $\lambda$ on the same order as the travel contribution.

\begin{figure}[t]
\centering
\includegraphics[width=\columnwidth]{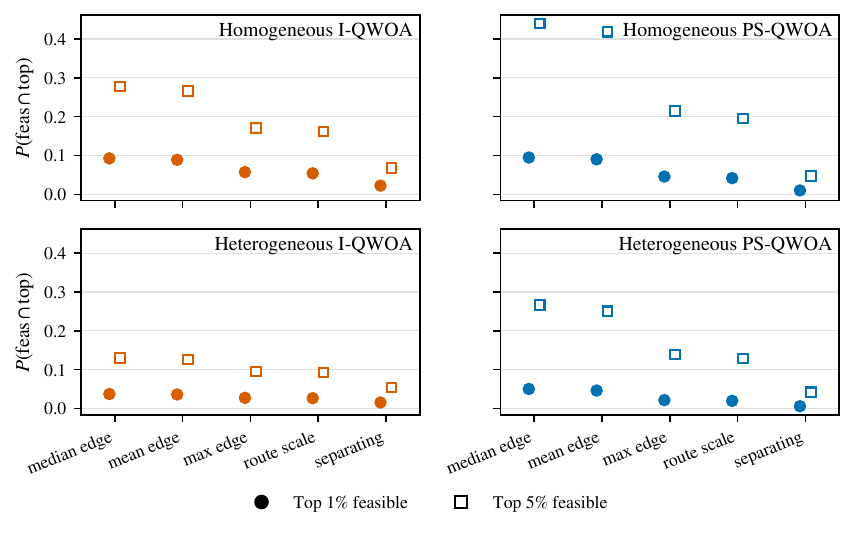}
\caption{Penalty-heuristic comparison at depth $p=8$ on homogeneous and heterogeneous $(n,K)=(5,3)$ CVRP instances for PS-QWOA and I-QWOA showing the average measurement probability of solutions in the top 1\% and 5\% of the feasible solutions.}
\label{fig:penalty_ranking}
\end{figure}

\begin{table}
\centering
\caption{Averaged performance at $p=8$ on the $(n,K)=(5,3)$ instances, across the two algorithms and two variants. The $\lambda$ column gives the numerical value of each rule on the homogeneous seed; remaining columns are as defined in \cref{fig:penalty_ranking}, with $P_{\mathrm{opt}\mid\mathrm{feas}}=P_{\mathrm{opt}}/P_{\mathrm{feas}}$ and $r_{\mathrm{feas}}$ as in \cref{eq:r_feas}.}
\label{tab:penalty_results}
\begin{tabular}{lrrrrr}
\toprule
Rule & $\lambda$ & $P_{\mathrm{feas}}$ & $P_{\mathrm{opt}\mid\mathrm{feas}}$ & $P_{\mathrm{opt}}$ & $r_{\mathrm{feas}}$ \\
\midrule
$\lambda_{\text{sep}}$   & 453.0 & \textbf{0.992} & 0.014 & 0.012 & 1.187 \\
$\lambda_{\max}$         &  25.1 & 0.982 & 0.039 & 0.037 & 1.120 \\
$\lambda_{\text{route}}$ &  27.2 & 0.984 & 0.036 & 0.035 & 1.126 \\
$\lambda_{\text{mean}}$  &  12.7 & 0.954 & 0.068 & 0.065 & 1.073 \\
$\lambda_{\text{med}}$   &  11.8 & 0.949 & \textbf{0.072} & \textbf{0.069} & \textbf{1.069} \\
\bottomrule
\end{tabular}
\end{table}

\FloatBarrier

Two trends emerge (\cref{fig:penalty_ranking}, \cref{tab:penalty_results}). First, each candidate concentrates $\ge 0.95$ of the measurement probability on the feasible subspace, but the magnitude of the penalty has a significant effect on convergence to low-cost solutions within that space. At roughly forty times the mean edge cost, $\lambda_{\text{sep}}$ drops $P_{\mathrm{opt}\mid\mathrm{feas}}$ to $\approx 0.01$ and raises $r_{\mathrm{feas}}$ to $\approx 1.19$, whereas $\lambda_{\text{mean}}$ and $\lambda_{\text{med}}$ give a five-fold increase in $P_{\mathrm{opt}}$ over $\lambda_{\text{sep}}$. The intermediate rules $\lambda_{\max}$ and $\lambda_{\text{route}}$ interpolate between these extremes. We adopt $\lambda_{\text{mean}}$ \eqref{eq:lambda_mean} as the default rule for \cref{sec:numerical}.

\end{appendix}

\end{document}

%% file: figures/walk_generator_graphs.tex
\begin{figure*}[t]
  \centering
  \begingroup
  \definecolor{walkassign}{RGB}{47,108,158}
  \definecolor{transswapA}{RGB}{230,159,0}
  \definecolor{transswapB}{RGB}{0,158,115}
  \definecolor{transswapC}{RGB}{204,121,167}
  \pgfdeclarelayer{bg}
  \pgfsetlayers{bg,main}
  \tikzset{
    panelbox/.style={rounded corners=4pt, draw=white, line width=0.45pt},
    walknode/.style={circle, draw=black!70, fill=white, line width=0.55pt, inner sep=1.2pt},
    assignedge/.style={draw=walkassign, line width=0.9pt},
    trans12edge/.style={draw=transswapA, line width=0.8pt},
    trans13edge/.style={draw=transswapB, line width=0.8pt},
    trans23edge/.style={draw=transswapC, line width=0.95pt},
    product12edge/.style={draw=transswapA, line width=0.32pt, opacity=0.5},
    product13edge/.style={draw=transswapB, line width=0.32pt, opacity=0.5},
    product23edge/.style={draw=transswapC, line width=0.46pt, opacity=0.6,
      preaction={draw=white, line width=0.95pt, opacity=0.5}},
    paneltitle/.style={font=\bfseries\small, text=black!85},
    panelsub/.style={font=\scriptsize\itshape, text=black!60},
    nodelabel/.style={font=\scriptsize, text=black!85}
  }
  \def\drawcube#1#2#3#4#5{%
    \begin{scope}[shift={#2}, scale=#3]
      \coordinate (#1111) at (0.00,0.00);
      \coordinate (#1211) at (0.60,0.00);
      \coordinate (#1121) at (0.00,0.60);
      \coordinate (#1221) at (0.60,0.60);
      \coordinate (#1112) at (0.28,0.22);
      \coordinate (#1212) at (0.88,0.22);
      \coordinate (#1122) at (0.28,0.82);
      \coordinate (#1222) at (0.88,0.82);

      \draw[assignedge] (#1111) -- (#1211);
      \draw[assignedge] (#1111) -- (#1121);
      \draw[assignedge] (#1211) -- (#1221);
      \draw[assignedge] (#1121) -- (#1221);
      \draw[assignedge] (#1112) -- (#1212);
      \draw[assignedge] (#1112) -- (#1122);
      \draw[assignedge] (#1212) -- (#1222);
      \draw[assignedge] (#1122) -- (#1222);
      \draw[assignedge] (#1111) -- (#1112);
      \draw[assignedge] (#1211) -- (#1212);
      \draw[assignedge] (#1121) -- (#1122);
      \draw[assignedge] (#1221) -- (#1222);

      \node[walknode] at (#1111) {};
      \node[walknode] at (#1211) {};
      \node[walknode] at (#1121) {};
      \node[walknode] at (#1221) {};
      \node[walknode] at (#1112) {};
      \node[walknode] at (#1212) {};
      \node[walknode] at (#1122) {};
      \node[walknode] at (#1222) {};

      \ifnum#5=1\relax
        \node[nodelabel, anchor=north east] at ($(#1111)+(-0.03,-0.03)$) {$111$};
        \node[nodelabel, anchor=north west] at ($(#1211)+(0.03,-0.03)$) {$211$};
        \node[nodelabel, anchor=south east] at ($(#1121)+(-0.05,0.02)$) {$121$};
        \node[nodelabel, anchor=south west] at ($(#1221)+(-0.30,-0.18)$) {$221$};
        \node[nodelabel, anchor=north east] at ($(#1112)+(0.29,0.18)$) {$112$};
        \node[nodelabel, anchor=north west] at ($(#1212)+(0.06,-0.03)$) {$212$};
        \node[nodelabel, anchor=south east] at ($(#1122)+(-0.06,0.04)$) {$122$};
        \node[nodelabel, anchor=south west] at ($(#1222)+(0.06,0.04)$) {$222$};
      \fi

      \ifx\relax#4\relax\else
        \node[font=\scriptsize, text=black!80] at (0.44,-0.34) {$#4$};
      \fi
    \end{scope}
  }

  \resizebox{\linewidth}{!}{%
    \begin{tikzpicture}[x=1cm, y=1cm]
      \draw[panelbox] (0.00,0.00) rectangle (4.95,5.85);
      \draw[panelbox] (5.25,0.00) rectangle (9.30,5.85);
      \draw[panelbox] (9.60,0.00) rectangle (17.60,5.85);

      \node[paneltitle] at (2.48,5.42) {$H(3,2)$};
      \node[panelsub] at (2.48,5.03) {fixed $\pi = 123$};
      \drawcube{A}{(1.36,1.25)}{2.20}{}{1}

      \node[paneltitle] at (7.28,5.42) {$S_3$};
      \node[panelsub] at (7.28,5.03) {fixed $a = 112$};
      \coordinate (P123) at (6.20,3.95);
      \coordinate (P231) at (6.20,2.68);
      \coordinate (P312) at (6.20,1.41);
      \coordinate (P132) at (8.35,3.95);
      \coordinate (P213) at (8.35,2.68);
      \coordinate (P321) at (8.35,1.41);
      \draw[trans23edge] (P123) -- (P132);
      \draw[trans12edge] (P123) -- (P213);
      \draw[trans13edge] (P123) -- (P321);
      \draw[trans13edge] (P231) -- (P132);
      \draw[trans23edge] (P231) -- (P213);
      \draw[trans12edge] (P231) -- (P321);
      \draw[trans12edge] (P312) -- (P132);
      \draw[trans13edge] (P312) -- (P213);
      \draw[trans23edge] (P312) -- (P321);
      \node[walknode] at (P123) {};
      \node[walknode] at (P231) {};
      \node[walknode] at (P312) {};
      \node[walknode] at (P132) {};
      \node[walknode] at (P213) {};
      \node[walknode] at (P321) {};
      \node[nodelabel] at ($(P123)+(-0.52,0.00)$) {$123$};
      \node[nodelabel] at ($(P231)+(-0.52,0.00)$) {$231$};
      \node[nodelabel] at ($(P312)+(-0.52,0.00)$) {$312$};
      \node[nodelabel] at ($(P132)+(0.52,0.00)$) {$132$};
      \node[nodelabel] at ($(P213)+(0.52,0.00)$) {$213$};
      \node[nodelabel] at ($(P321)+(0.52,0.00)$) {$321$};

        \node[paneltitle] at (13.60,5.42) {$H(3,2)\,\square\,S_3$};
        \node[panelsub] at (13.60,5.03) {on $(\pi,a)$};
      \draw[assignedge] (10.35,4.58) -- (10.92,4.58);
      \node[panelsub, anchor=west] at (11.04,4.58) {assignment edge};
      \draw[trans12edge] (13.18,4.58) -- (13.52,4.58);
      \node[panelsub, anchor=west] at (13.60,4.58) {$(12)$};
      \draw[trans13edge] (14.42,4.58) -- (14.76,4.58);
      \node[panelsub, anchor=west] at (14.84,4.58) {$(13)$};
      \draw[trans23edge] (15.66,4.58) -- (16.00,4.58);
      \node[panelsub, anchor=west] at (16.08,4.58) {$(23)$};

      \begin{scope}[shift={(0.75,0)}]
        \drawcube{c123v}{(10.20,3.46)}{0.92}{123}{0}
        \drawcube{c231v}{(10.20,2.03)}{0.92}{231}{0}
        \drawcube{c312v}{(10.20,0.60)}{0.92}{312}{0}
        \drawcube{c132v}{(14.32,3.46)}{0.92}{132}{0}
        \drawcube{c213v}{(14.32,2.03)}{0.92}{213}{0}
        \drawcube{c321v}{(14.32,0.60)}{0.92}{321}{0}

        \begin{pgfonlayer}{bg}
          \foreach \vertex in {111,211,121,221,112,212,122,222} {
            \draw[product13edge] (c123v\vertex) to[out=-7,in=-173] (c321v\vertex);
            \draw[product13edge] (c231v\vertex) to[out=-7,in=-173] (c132v\vertex);
            \draw[product13edge] (c312v\vertex) to[out=-7,in=-173] (c213v\vertex);

            \draw[product12edge] (c123v\vertex) to[out=7,in=173] (c213v\vertex);
            \draw[product12edge] (c231v\vertex) to[out=7,in=173] (c321v\vertex);
            \draw[product12edge] (c312v\vertex) to[out=7,in=173] (c132v\vertex);

            \draw[product23edge] (c123v\vertex) -- (c132v\vertex);
            \draw[product23edge] (c231v\vertex) -- (c213v\vertex);
            \draw[product23edge] (c312v\vertex) -- (c321v\vertex);
          }
        \end{pgfonlayer}
      \end{scope}
    \end{tikzpicture}%
  }
  \caption{PS-QWOA walk-generator graphs for $n=3$ and $K=2$. \textbf{Left:} with $\pi=(1,2,3)$ fixed, $\mathcal G_a$ produces the assignment graph $H(3,2)$ on the eight possible assignments, denoted as $a_1a_2a_3$. \textbf{Centre:} with $a=(1,1,2)$ fixed, $\mathcal G_\pi$ produces the transposition graph on $S_3$. Colours distinguishing the three transpositions $(12)$, $(13)$, and $(23)$. \textbf{Right:} the composite generator $\mathcal G_a + \mathcal G_\pi$ is a walk over the graph $H(3,2)\square S_3$, where $\square$ denotes the graph Cartesian product.}
  \label{fig:walk_generator_graphs_n3k2}
  \endgroup
\end{figure*}

%% file: figures/symmetric_subspace_graphs.tex
\begin{figure*}[t]
  \centering
  \begingroup
  \definecolor{walkassign}{RGB}{47,108,158}
  \definecolor{transswapA}{RGB}{230,159,0}
  \definecolor{transswapB}{RGB}{0,158,115}
  \definecolor{transswapC}{RGB}{204,121,167}
  \pgfdeclarelayer{bg}
  \pgfsetlayers{bg,main}
  \tikzset{
    panelbox/.style={rounded corners=4pt, draw=white, line width=0.45pt},
    walknode/.style={circle, draw=black!70, fill=white, line width=0.55pt, inner sep=1.2pt},
    assignedge/.style={draw=walkassign, line width=0.9pt},
    trans12edge/.style={draw=transswapA, line width=0.8pt},
    trans13edge/.style={draw=transswapB, line width=0.8pt},
    trans23edge/.style={draw=transswapC, line width=0.95pt},
    product12edge/.style={draw=transswapA, line width=0.32pt, opacity=0.5},
    product13edge/.style={draw=transswapB, line width=0.32pt, opacity=0.5},
    product23edge/.style={draw=transswapC, line width=0.46pt, opacity=0.6,
      preaction={draw=white, line width=0.95pt, opacity=0.5}},
    paneltitle/.style={font=\bfseries\small, text=black!85},
    panelsub/.style={font=\scriptsize\itshape, text=black!60},
    nodelabel/.style={font=\scriptsize, text=black!85},
    weightlabel/.style={font=\scriptsize\itshape, text=black!55}
  }
  \def\drawkfour#1#2#3#4#5{%
    \begin{scope}[shift={#3}, scale=#4]
      \coordinate (#1v111) at (0.00,0.00);
      \coordinate (#1v112) at (0.72,0.00);
      \coordinate (#1v121) at (0.00,0.72);
      \coordinate (#1v122) at (0.72,0.72);

      \draw[assignedge] (#1v111) -- (#1v112);
      \draw[assignedge] (#1v111) -- (#1v121);
      \draw[assignedge] (#1v111) -- (#1v122);
      \draw[assignedge] (#1v112) -- (#1v121);
      \draw[assignedge] (#1v112) -- (#1v122);
      \draw[assignedge] (#1v121) -- (#1v122);

      \node[walknode] at (#1v111) {};
      \node[walknode] at (#1v112) {};
      \node[walknode] at (#1v121) {};
      \node[walknode] at (#1v122) {};

      \ifnum#5=1\relax
        \node[nodelabel] at ($(#1v111)+(-0.16,-0.12)$) {$111$};
        \node[nodelabel] at ($(#1v112)+(0.16,-0.12)$) {$112$};
        \node[nodelabel] at ($(#1v121)+(-0.16,0.12)$) {$121$};
        \node[nodelabel] at ($(#1v122)+(0.16,0.12)$) {$122$};
      \fi

      \ifx\relax#2\relax\else
        \node[font=\scriptsize, text=black!80] at (0.36,-0.48) {$#2$};
      \fi
    \end{scope}
  }

  \makebox[\linewidth][c]{%
    \resizebox{0.79\linewidth}{!}{%
      \begin{tikzpicture}[x=1cm, y=1cm]
      \draw[panelbox] (0.00,0.00) rectangle (4.45,5.60);
      \draw[panelbox] (4.75,0.00) rectangle (13.25,5.60);

      \node[paneltitle] at (2.23,5.20) {$K_4$};
      \node[panelsub] at (2.23,4.82) {fixed $\pi = 123$};
      \drawkfour{kfourA}{}{(1.18,1.75)}{2.00}{1}

      \node[paneltitle] at (9.00,5.20) {$K_4\,\square\,S_3$};
      \node[panelsub] at (9.00,4.82) {on $(\pi,s)$};
      \draw[assignedge] (5.51,4.38) -- (6.09,4.38);
      \node[panelsub, anchor=west] at (6.21,4.38) {assignment edge};
      \draw[trans12edge] (8.65,4.38) -- (8.97,4.38);
      \node[panelsub, anchor=west] at (9.05,4.38) {$(12)$};
      \draw[trans13edge] (9.81,4.38) -- (10.13,4.38);
      \node[panelsub, anchor=west] at (10.21,4.38) {$(13)$};
      \draw[trans23edge] (10.97,4.38) -- (11.29,4.38);
      \node[panelsub, anchor=west] at (11.37,4.38) {$(23)$};

      \begin{scope}[shift={(0.75,0.20)}]
        \drawkfour{k123}{123}{(5.50,3.05)}{0.86}{0}
        \drawkfour{k231}{231}{(5.50,1.71)}{0.86}{0}
        \drawkfour{k312}{312}{(5.50,0.37)}{0.86}{0}
        \drawkfour{k132}{132}{(10.20,3.05)}{0.86}{0}
        \drawkfour{k213}{213}{(10.20,1.71)}{0.86}{0}
        \drawkfour{k321}{321}{(10.20,0.37)}{0.86}{0}

        \begin{pgfonlayer}{bg}
          \foreach \vertex in {111,112,121,122} {
            \draw[product13edge] (k123v\vertex) to[out=-7,in=-173] (k321v\vertex);
            \draw[product13edge] (k231v\vertex) to[out=-7,in=-173] (k132v\vertex);
            \draw[product13edge] (k312v\vertex) to[out=-7,in=-173] (k213v\vertex);

            \draw[product12edge] (k123v\vertex) to[out=7,in=173] (k213v\vertex);
            \draw[product12edge] (k231v\vertex) to[out=7,in=173] (k321v\vertex);
            \draw[product12edge] (k312v\vertex) to[out=7,in=173] (k132v\vertex);

            \draw[product23edge] (k123v\vertex) -- (k132v\vertex);
            \draw[product23edge] (k231v\vertex) -- (k213v\vertex);
            \draw[product23edge] (k312v\vertex) -- (k321v\vertex);
          }
        \end{pgfonlayer}
      \end{scope}
      \end{tikzpicture}%
    }%
  }
  \caption{Reduced graph structure in the relabelling-symmetric subspace for $n=3$ and $K=2$. Left: for fixed $\pi$, the quotient of the assignment graph by global vehicle relabelling is the reduced assignment graph on the four canonical restricted-growth representatives $111$, $112$, $121$, and $122$; this graph is isomorphic to $K_4$, and every edge has weight $1/3$. Right: the composite reduced graph acts on the Cartesian product $K_4\square S_3$; each small reduced-assignment fibre corresponds to a fixed permutation, while the coloured inter-fibre edges show the three transposition classes lifted to the reduced subspace. The permutation factor is the same transposition graph on $S_3$ shown in the centre panel of \cref{fig:walk_generator_graphs_n3k2}.}
  \label{fig:symmetric_subspace_graphs_n3k2}
  \endgroup
\end{figure*}